\documentclass[%
 reprint,
superscriptaddress,
 amsmath,amssymb,
 aps,
 pra, 
longbibliography
]{revtex4-1}

\usepackage{graphicx}
\usepackage{float}
\usepackage{amsmath}
\usepackage{color}
\usepackage{dcolumn}
\usepackage{bm}
\usepackage{hyperref}
\usepackage{braket}
\usepackage{xcolor}
\usepackage{enumerate}
\usepackage{comment}
\usepackage{dsfont}
\usepackage{wrapfig}
\usepackage{pdfpages}
\usepackage{ulem}
\usepackage{multirow}

\makeatletter
\AtBeginDocument{\let\LS@rot\@undefined}
\makeatother

\begin{document}

\title{Fusion channels of non-Abelian anyons from angular-momentum and density-profile measurements}

\author{E. Macaluso}
\affiliation{%
INO-CNR BEC Center and Dipartimento di Fisica, Universit$\grave{a}$ di Trento, 38123 Trento, Italy
}%
\author{T. Comparin}
\affiliation{%
INO-CNR BEC Center and Dipartimento di Fisica, Universit$\grave{a}$ di Trento, 38123 Trento, Italy
}%
\author{L. Mazza}%
\affiliation{%
LPTMS, CNRS, Universit$\acute{e}$ Paris-Sud, Universit$\acute{e}$ Paris-Saclay, 91405 Orsay, France
}%

\author{I. Carusotto}%
\affiliation{%
INO-CNR BEC Center and Dipartimento di Fisica, Universit$\grave{a}$ di Trento, 38123 Trento, Italy
}%

\date{\today}

\begin{abstract}
We present a method to characterize non-Abelian anyons that is based only on static measurements and that does not rely on any form of interference. For geometries where the anyonic statistics can be revealed by rigid rotations of the anyons, we link this property to the angular momentum of the initial state. We test our method on the paradigmatic example of the Moore-Read state, that is known to support excitations with non-Abelian statistics of Ising type. As an example, we reveal the presence of different fusion channels for two such excitations, a defining feature of non-Abelian anyons. This is obtained by measuring density-profile properties, like the mean square radius of the system or the depletion generated by the anyons. Our study paves the way to novel methods for characterizing non-Abelian anyons, both in the experimental and theoretical domains.
\end{abstract}

\maketitle

{\it Introduction.---}
The standard classification of particles into bosons and fermions breaks down in two spatial dimensions, where exotic objects known as \textit{anyons} can  exist~\cite{Wilczek_PRL.49.957, Halperin_PRL.52.1583, Arovas_Wilczek_PRL.53.722, Leinaas_Myrheim_Nuovo_Cimento, Wu_PRL.52.2103, Stern_AoP}.
The key concepts for defining the statistics of anyons are the adiabatic motion of one anyon around another, hereafter the \textit{braiding}, and the adiabatic \textit{exchange} of the anyons positions~\cite{Tong_notes}.
Anyons can be characterized by
merging two of them, and the properties of the new composite object depend on the fusion rules of the original anyons. 
When there is the possibility of fusing in more than one way, anyons can be non-Abelian~\cite{Moore_NPB.360.2.362, ReadGreen_PRB.61.10267, Ivanov_PRL.86.268, Read_Rezayi_PRB.59.8084}:
they are the heart of topological quantum computation~\cite{Nayak_RMP.80.1083}, and their experimental realization is thus highly desired.
Several existing platforms are expected to host them 
as emergent quasi-particles, but the unambiguous experimental demonstration of their properties is still the matter of an intense debate~\cite{Willett_PNAS.106.22, Mourik_Science.336.6084}.

In the last twenty years, several works addressed the problem of extracting the properties of the anyons hosted by the ground states of a given Hamiltonian.
The simplest approach relies on explicitly following the ground-state evolution when anyons are exchanged~\cite{ParedesZoller_PRL.87.010402, Tserkovnyak_PRL.90.016802, Baraban_Simon_numerical_analysis_MR_qh_wfs, Wu_PRL.113.116801, Nielsen_PRB.91.041106}.
Within other approaches, the analytical study of paradigmatic wave functions has also clarified important issues about the statistics of excitations~\cite{NayakWilczek_NuclPhysB479.3_1996, Nayak_plasma_Ising-type_FQH, Ivanov_PRL.86.268}.
On the experimental side, interferometric schemes have been proposed to compare the state before and after the adiabatic time evolution~\cite{Halperin_Rosenow_PRB.83.155440, Campagnano_Gefen_PRL.109.106802, ParedesZoller_PRL.87.010402, DasSarma_Nayak_PRL.94.166802, Stern_Halperin_PRL.96.016802, Bonderson_Shtengel_PRL.96.016803}, but none of them has produced unambiguous results~\cite{Camino_Goldman_PRB.72.075342, Rosenow_Halperin_PRL.98.106801}.

\begin{figure}[tb]
\centering
\includegraphics[width=0.48\textwidth]{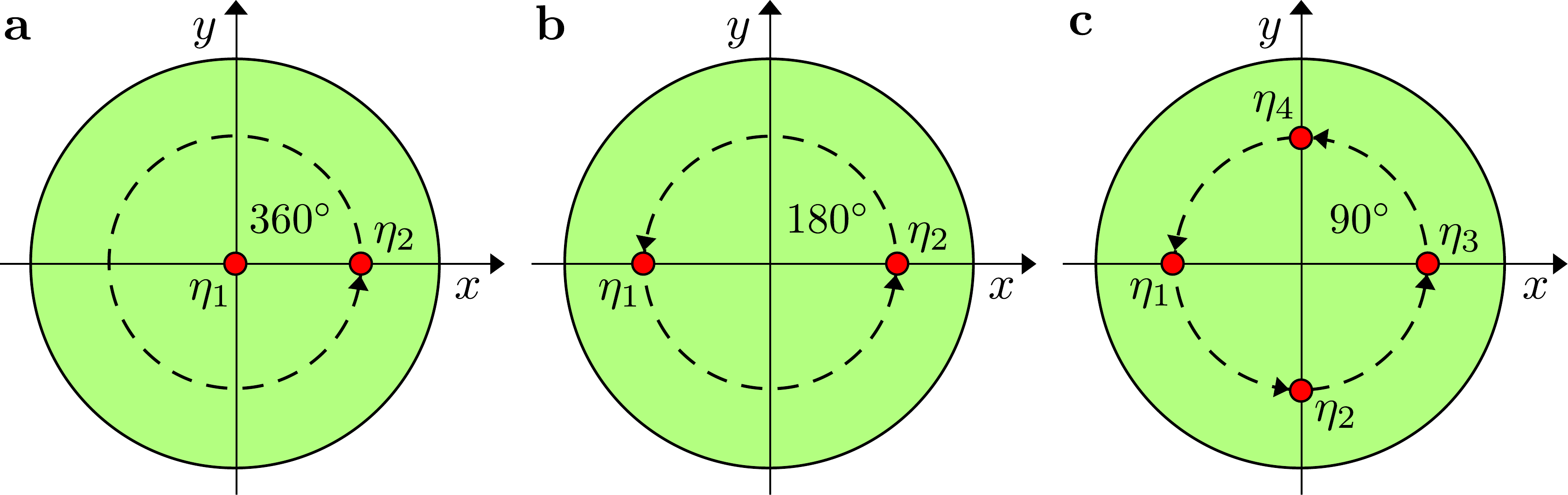}
\caption{\textbf{a}-\textbf{c}, Rigid rotations of two anyons (panels {\bf a} and {\bf b}) or four anyons (panels {\bf c}). Rotation angles are such that the set of anyonic coordinates (red dots) remains the same.}
\label{fig:braiding_rigid_rotations}
\end{figure}

We propose a method to characterize non-Abelian anyons:
By considering geometries where the anyonic statistics can be revealed through rigid rotations of the anyons (see Fig.~\ref{fig:braiding_rigid_rotations}), we relate their statistical phase to the angular momentum and to the density profile of the system.
This protocol allows one to identify the existence of different fusion channels, a defining property of non-Abelian anyons,
with remarkable experimental simplicity in the context of ultracold atoms~\cite{Pitaevskii_Stringari_2016, Cooper_etal_TopoBands_RMP} and photons~\cite{ Carusotto_Ciuti_RMP.85.299, Ozawa_etal_TopoPhoto_RMP}.
Moreover, our study represents a powerful theoretical tool to inspect excitations with unknown statistics, going beyond the observation of multiple fusion channels.
As a showcase study, we discuss our method for the case of the Moore-Read (MR) state~\cite{Moore_NPB.360.2.362}, and outline an experimental procedure for computing the statistical phases of its quasiholes.

{\it Rigid rotations of the anyons.---}
We consider a two-dimensional (2D) system of $N$ particles (bosons or fermions) supporting anyonic excitations.
The Hamiltonian $\hat H_1$ is a function of particle positions and momenta, as well as of time.
We use the complex coordinate notation $z_j = x_j + i y_j$ for the position of the $j$-th particle.
The time dependence of $\hat H_1 (\partial_{z_j}, \partial_{\bar z_j},z_j, \bar z_j;  t)$ is only due to a set of parameters $\eta_\mu(t)$ defining the centers of some external local potentials $V_\mathrm{ext}(z, \eta_\mu(t))$.
These potentials typically couple with the particle density, creating and pinning the anyons at positions $\eta_\mu(t)$ \cite{ParedesZoller_PRL.87.010402, MacalusoCarusotto_PRA.98.013605, Wan_Rezayi_PRL.97.256804, Wan_Yang_PRB.77.165316, Prodan_Haldane_PRB.80.115121}.

To reveal the anyonic statistics,
one option is to braid the anyons through rigid rotations of the pinning-potential coordinates (see Fig.~\ref{fig:braiding_rigid_rotations}). These transformations are defined as
\begin{equation}
\eta_{\mu} (t) = \eta_{\mu} (0) e^{i \theta(t)},
\qquad \theta(t) = \frac {t}{T} \theta_f ,
\label{Eq:Anyon:Rotation}
\end{equation}
where $\theta_f$ is the final rotation angle and $T$ is the time duration of the process.

Since we consider rigid rotations, we can study the problem in the reference frame $R_2$ co-rotating with the anyons, rather than using the laboratory reference frame $R_1$.
We assume that $V_\mathrm{ext}(z,\eta_\mu(t))$ is a function of the distance $|z - \eta_\mu(t)|$ between particles and anyons, and that the remaining terms in $\hat{H}_1$ are rotationally invariant.
Under these assumptions, the generator of the time evolution in $R_2$ in the time span $[0,T]$ reads~\cite{Pitaevskii_Stringari_2016}:
\begin{equation}
 \hat H_2 (\partial_{z_j}, \partial_{\bar z_j},z_j, \bar z_j;  t) =
 \hat H_1 (\partial_{z_j}, \partial_{\bar z_j},z_j, \bar z_j;  t=0)
 - \frac{\theta_f}{T} \hat L_z,
\end{equation}
which is manifestly time-independent. 
The first term on the right-hand side is the initial Hamiltonian in $R_1$, while the second one describes the effect of the rotation. Being interested in an adiabatic process, we consider $T \to \infty$. The rotation term is then a small contribution and can be treated perturbatively.

To describe the dynamics in $R_2$, we consider an initial state $\ket{\Psi_0}$ belonging to the $m$-fold degenerate ground-state manifold $ \mathcal{H}_{E_0}$, spanned by the basis $\lbrace\ket{\psi_{\alpha}}\rbrace_{\alpha=1,\dots,m}$, with $\hat{H}_1(t=0)\ket{\psi_{\alpha}}=E_0\ket{\psi_{\alpha}}$ and $\braket{\psi_{\alpha}|\psi_{\beta}} = \delta_{\alpha\beta}$.
If the dynamics is slow enough, we can use the adiabatic theorem to state that the dynamics is restricted to $\mathcal{H}_{E_0}$ (an explicit proof is in~\cite{SuppMat}), and make the following ansatz:
\begin{equation}
\ket{\Psi_2(t)} =
e^{- i E_0 t / \hbar} \sum_{\alpha = 1}^m \gamma_{\alpha}(t) \ket{\psi_{\alpha}},
\quad  \gamma_\alpha(0) = \braket{\psi_{\alpha} | \Psi_0}. \label{Eq:State:2:Ansatz}
\end{equation}
By applying the Schr\"odinger equation, we recover the time-evolution equation of the $\gamma_\alpha$'s:
\begin{equation}
 i \hbar \frac{\mathrm d \gamma_{\alpha}(t)}{\mathrm d t} =
 -\frac{\theta_f}{T}\sum_{\beta=1}^m 
 \mathcal{L}_{\alpha\beta}\,  \gamma_{\beta} (t),
 \label{Eq:Diff:gamma}
\end{equation}
where $\mathcal{L}_{\alpha\beta} = \braket{\psi_{\alpha} | \hat{L}_z | \psi_{\beta}}$ is the angular momentum restricted to $\mathcal{H}_{E_{0}}$.
The solution reads
\begin{equation}
 \ket{\Psi_2(T)} = 
 e^{- i \hat{H}_2 T / \hbar}  \ket{\Psi_0}
 =
 e^{- i E_0 T / \hbar}\,
 e^{i \theta_f \mathcal L / \hbar}
 \ket{\Psi_0},
 \label{Eq:Psi2}
\end{equation}
in terms of the matrix exponential $\exp \left[i \theta_f \mathcal L / \hbar \right]$.

To find the state $\ket{\Psi_1(T)}$ in the laboratory frame, we need to rotate $\ket{\Psi_2(T)}$ by an angle $\theta_f$:
\begin{equation}
\begin{aligned}
\ket{\Psi_1(T)}
&=
e^{- i \theta_f \hat{ L}_z / \hbar}  \ket{\Psi_2(T)}
\\ &=
e^{- i E_0 T / \hbar}\,
e^{- i \theta_f \hat{ L}_z / \hbar}\,
e^{i \theta_f \mathcal L / \hbar}  \ket{\Psi_0}.
\end{aligned}
\label{Eq:Psi1}
\end{equation}
The state in equation~\eqref{Eq:Psi1} is the exact result for an adiabatic braiding process performed through a rigid rotation of all anyons by an angle $\theta_f$.
We recognize a dynamical phase proportional to $T$, that is unessential to the discussion of non-Abelian statistics and therefore neglected from now on.
The remaining geometric contribution is the product of two unitary matrices:
$\mathcal{B}$, with matrix elements $\mathcal{B}_{\alpha\beta} = \bra{\psi_{\alpha}} e^{- i \theta_f \hat L_z /\hbar} \ket{\psi_{\beta}}$, and $\mathcal{U}_\mathrm{B} \equiv e^{i \theta_f \mathcal L / \hbar}$, which is the Berry matrix of the adiabatic process under study, once one makes a suitable choice of the basis states for each angle $\theta(t)$~\cite{SuppMat}.

To guarantee that the ground-state manifold is $\mathcal{H}_{E_0}$ at both times~\cite{Nayak_RMP.80.1083}, the angle $\theta_f$ must be such that $\hat{H}_1(t)$ is the same at times $t=0$ and $T$.
Depending on the anyon positions, this constraint can be satisfied even for rotation angles which are not multiple of $2\pi$ [see Fig.~\ref{fig:braiding_rigid_rotations} \textbf{b}-\textbf{c}].
When $\theta_f = 2 \pi k$, with $k$ integer, $\mathcal{B}$ is trivially the identity matrix. In this case, $\mathcal{U}_\mathrm{B}$ encodes the full geometrical contribution to the time evolution, made up of both topological and non-topological parts.
We stress that $\mathcal{U}_\mathrm{B}$ only depends on measurable properties of the ground-state manifold at the initial time, namely the angular-momentum matrix elements.
Therefore no actual time evolution is needed to measure it, which constitutes an undeniable experimental advantage.
The case of $\theta_f \neq 2 \pi k$ is relevant in the theoretical context, where --in contrast with experimental studies-- nothing precludes the extraction of $\mathcal{B}$ [see example in Ref.~\cite{SuppMat}]. A comprehensive analysis of this case is left for a future work.

\begin{figure*}[htb]
\centering
\includegraphics[width=0.99\textwidth]{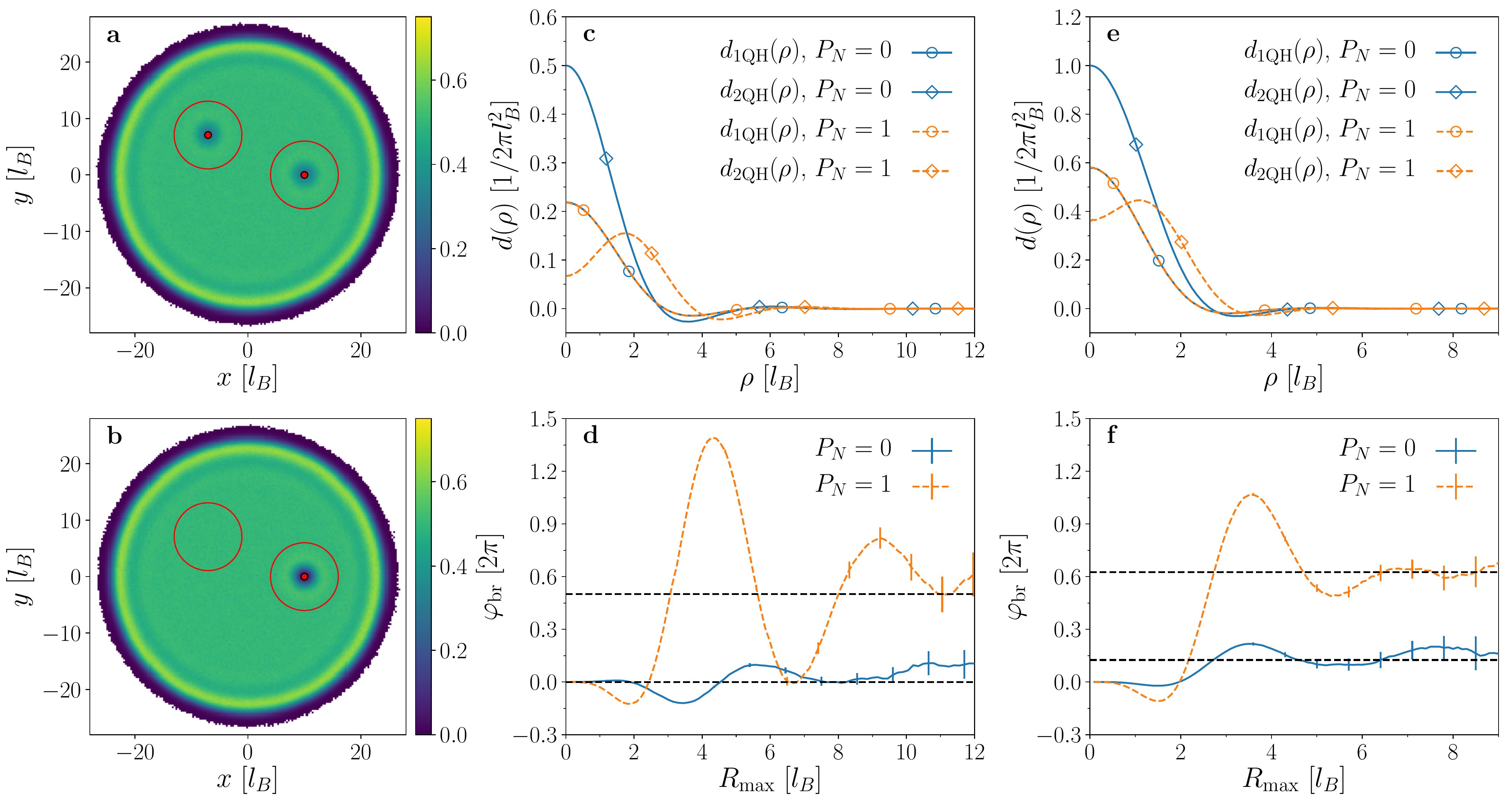}
\caption{\textbf{a}, 2D density profile of the $N=150$ $M=2$ Moore-Read state with quasiholes at positions $\eta_{1} = 10 \, l_{\text{B}}$ and $\eta_{2} = 10 e^{i3\pi/2} \, l_{\text{B}}$ (red dots).
\textbf{b}, 2D density profile of the $N=150$ $M=2$ Moore-Read state with quasiholes at positions $\eta_{1} = \eta_{2} = 10 \, l_{\text{B}}$. Red circles give a pictorial representation of the regions $A_{1}$ and $A_{2}$ where the 2D densities depicted in \textbf{a} and \textbf{b} are different.
\textbf{c}, Radial profile of the density depletions caused by a single quasihole at $\eta_1 = 0$ (circles) and two quasiholes on top of each other at $\eta_1 = \eta_2 = 0$ (diamonds), for even (blue solid lines) and odd (orange dashed lines) parity of $N$, at filling $\nu = 1/M = 1/2$. We consider $N=200$ or $N=199$.
\textbf{d}, Quasihole braiding phase evaluated with equation~\eqref{eq:phi_br_from_depletions} as a function of the cutoff radius $R_{\text{max}}$, for both $P_{N}=0$ (blue solid line) and $P_{N}=1$ (orange dashed line) in the $M=2$ fermionic case. Black dashed lines denote the predictions for $\varphi_{\text{br}}$ [see equation~\eqref{eq:phi_br_MR_QHs}]. \textbf{e}, \textbf{f}, Same as \textbf{c}, \textbf{d} for the $M=1$ bosonic case.}
\label{fig:phi_br_depletions}
\end{figure*}

{\it Moore-Read state and its quasihole excitations.---}
We now consider the MR state, which is described by the wave function~\cite{Moore_NPB.360.2.362}
\begin{equation}
    \Psi (\{z_j\}) = \text{Pf}(W) \prod_{i<j} \left(z_{i} - z_{j} \right)^{M} e^{- \sum_{i} |z_{i}|^{2} / 4 l^{2}_{B} },
    \label{eq:MR_wf}
\end{equation}
where $l_{\text{B}}$ is the magnetic length.
$\text{Pf}(W)$ denotes the Pfaffian of the $N\times N$ anti-symmetric matrix $W$, with $W_{ij} = 1 / (z_i - z_j)$ for $i\neq j$. For even (odd) values of the positive integer $M$, this wave function represents a fermionic (bosonic) FQH state at filling $\nu = 1/M$, which belongs to the lowest Landau level (LLL)~\cite{Tong_notes}.
This state is the ground state for 2D charged particles, in the presence of a transverse magnetic field and of a specific three-body repulsion~\cite{Greiter_PRL.66.3205}, and it is believed to be in the same universality class of the FQH state observed at filling $\nu=5/2$ \cite{Greiter_NPB.374.567,  Morf_PRL.80.1505, Rezayi_Haldane_PRL.84.4685}.

In the presence of properly designed external potentials, the ground state may also host a specific number of localized anyonic excitations~\cite{Wan_Rezayi_PRL.97.256804, Wan_Yang_PRB.77.165316, Prodan_Haldane_PRB.80.115121}.
The quasihole (QH) excitations of the MR state obey non-Abelian statistics~\cite{Moore_NPB.360.2.362, NayakWilczek_NuclPhysB479.3_1996, Nayak_RMP.80.1083}.
In particular, they are Ising anyons with an additional Abelian contribution to their statistical phase, and they can fuse in two different ways.
For a given set of coordinates $(\eta_{1}, \dots, \eta_{2n})$ of $2n$ such QHs, there is a $2^{n-1}$-fold degenerate set of states~\cite{NayakWilczek_NuclPhysB479.3_1996}. 

In the following, we will consider the case $2n = 2$, for which the system is not degenerate.
In this case, the MR wave function $\Psi^{2\text{QH}}$ has the same form as in Eq.~\eqref{eq:MR_wf}; yet the anti-symmetric matrix $W$ depends on the even/odd parity $P_{N} = 0,1$ of the particle number $N$.
For $P_N = 0$, it is $N \times N$ and reads
\begin{equation}
W_{ij} = \dfrac{(\eta_1 - z_i)(\eta_2 - z_j) + (i \leftrightarrow j)}{z_i - z_j}
\quad \forall \, i \neq j.
\label{eq:MR_2QH}
\end{equation}
For $P_N = 1$, on the other hand, $W$ is a $(N+1)\times(N+1)$ matrix. The $N \times N$ upper-left block is defined as in Eq.~\eqref{eq:MR_2QH}, while the entries of the $(N+1)$-th row (column) are equal to $+1$ ($-1$)~\cite{SuppMat}.

The fusion channel of the two QHs depends on $P_N$~\cite{Nayak_plasma_Ising-type_FQH}.
As a consequence, the braiding of two MR QHs induces a phase $\varphi_{\rm br}$ that depends on $P_N$:
\begin{equation}
    \dfrac{\varphi_{\text{br}}}{2\pi} = \dfrac{1}{4M} - \dfrac{1}{8} + \dfrac{P_{N}}{2} .
    \label{eq:phi_br_MR_QHs}
\end{equation}
The dependence of the braiding phase $\varphi_\text{br}$ on $P_N$ is thus a direct indication of the non-Abelian statistics of QHs, because it indicates that the two QHs are in different fusion channels when $N$ is even or odd~\cite{Nayak_RMP.80.1083}.

{\it $\varphi_{\text{br}}$ from the mean square radius.---}
As previously mentioned, for a $2\pi$-rotation of the QHs, $\mathcal{B}$ is the identity matrix.
For the non-degenerate MR state with two QHs, the unitary transformation $\mathcal{U}(T)$ associated with this process reduces to the phase factor $\mathcal{U}_{\text{B}} = e^{i \varphi_{\text{B}}}$, where $\varphi_{\text{B}} = 2\pi \mathcal{L}/\hbar$ is the Berry phase. In this case, $\mathcal{L}$ is the expectation value of the angular-momentum operator over the initial state, $\langle \hat{L}_{z} \rangle$.

The Berry phase $\varphi_{\text{B}}$ has a non-topological contribution, which can be interpreted as an Aharonov-Bohm phase~\cite{SuppMat}.
Although this phase factor contains information on the QH fractional charge, we have to remove it to isolate the QH braiding phase $\varphi_{\text{br}}$.
To this purpose we consider the difference between the Berry phases for two particular states [see Fig.~\ref{fig:phi_br_depletions} \textbf{a} and \textbf{b}]:
\begin{equation}
    \dfrac{\varphi_{\text{br}}}{2\pi} = \dfrac{1}{\hbar} \left[\langle \hat{L}_{z} \rangle_{|\eta_{1}|=|\eta_{2}|} - \langle \hat{L}_{z} \rangle_{\eta_{1} = \eta_{2}} \right] .
    \label{eq:phi_br_Lz}
\end{equation}
The expectation value $\langle \hat{L}_{z} \rangle_{|\eta_{1}|=|\eta_{2}|}$ is taken on a state with QHs sufficiently far from each other, at positions $\eta_{1}$ and $\eta_{2}$ such that $|\eta_1| = |\eta_2|$.
On the other hand, $\langle \hat{L}_{z} \rangle_{\eta_{1}=\eta_{2}}$ is measured on the state with the two QHs on top of each other at $\eta_{1} = \eta_{2}$ [for details, see Ref.~\cite{SuppMat}].

The mean angular momentum of a state in the LLL is related to its mean square radius: $\langle \hat{L}_{z} \rangle / \hbar + N = N  \langle r^{2} \rangle / 2 l^{2}_{B}$~\cite{Ho_Mueller_PRL.89.050401, Umucalilar_PRL.120.230403}.
This simplifies equation~\eqref{eq:phi_br_Lz} which reads
\begin{equation}
    \dfrac{\varphi_{\text{br}}}{2\pi} = \dfrac{N}{2 l^{2}_{B}} \left[\langle r^{2} \rangle_{|\eta_{1}|=|\eta_{2}|} - \langle r^{2} \rangle_{\eta_{1} = \eta_{2}} \right] .
    \label{eq:phi_br_msr}
\end{equation}
Moreover, within the LLL approximation, the mean square radius of the cloud, and so $\varphi_{\text{br}}$, can be measured after time-of-flight expansion~\cite{Read_Cooper_PRA.68.035601, Umucalilar_PRL.120.230403}.

To validate equation~\eqref{eq:phi_br_msr}, we compute $\langle r^{2} \rangle$ through the Monte Carlo technique~\cite{SuppMat}.
Numerical results --reported in Table~\ref{tab:phi_br_2qhs} for both $M=2$ (fermionic case) and $M=1$ (bosonic case) and for different parities $P_N$ of the particle number $N$-- are fully compatible with equation~\eqref{eq:phi_br_MR_QHs}. This demonstrates that the existence of multiple fusion channels for the MR QHs can be experimentally probed without braiding them.

\begin{table}[h]
\centering
\begin{tabular}{c | c | c | c}
$M$		&	$P_{N}$				&
$\varphi_{\text{br}}^\mathrm{MC}$  $[2 \pi]$ &
$\varphi_{\text{br}}$  $[2 \pi]$ \\
\hline
\multirow{2}{*}{$2$ (fermions)} &	$0$ 		&	$0.05\pm0.06$ 	 &	$0$ \\
&	$1$ 		&	$0.49\pm0.07$    &	$0.5$ \\
\hline
\multirow{2}{*}{$1$ (bosons)} &	$0$ 		&	$0.13\pm0.04$	 &	$0.125$ \\
&	$1$ 		&	$0.59\pm0.04$	 &	$0.625$ \\
\end{tabular}
\caption{Quasihole braiding phase $\varphi_{\text{br}}^\mathrm{MC}$ obtained numerically via equation~\eqref{eq:phi_br_msr} (third column, with the Monte Carlo statistical uncertainty) and its prediction $\varphi_{\text{br}}$ in equation~\eqref{eq:phi_br_MR_QHs} (fourth column), for $M=2, 1$ and for different parities $P_{N}$ of the particle number $N=150$ and $N=149$.
For the $|\eta_1| = |\eta_2|$ term in equation~\eqref{eq:phi_br_msr}, we set $\eta_1 = - \eta_2$, which is the optimal configuration for a finite-size system.
For $M=2$ ($M=1$) case, $|\eta_1|/l_B$ is equal to 7.5 (6.5).
}
\label{tab:phi_br_2qhs}
\end{table}

{\it $\varphi_{\text{br}}$ from the quasihole density depletions.---}
Although the protocol suggested in equation~\eqref{eq:phi_br_msr} is already close to the current experimental capabilities, it requires the ability to pin QHs with high precision and the knowledge of the particle number.
Moreover, $\varphi_\text{br}$ is difficult to compute for large systems, since it is a $\mathcal{O}(1)$ number obtained as the difference between two $\mathcal{O}(N^2)$ quantities.
However, equation~\eqref{eq:phi_br_msr} can be recast in a form which does not depend neither on $N$ nor on the precise QH positions, as we prove in the following.
Due to the incompressibility of the FQH states~\cite{Tong_notes}, the densities of the configurations under study only differ in the regions $A_1$ and $A_2$ surrounding the QHs [see red circles in Fig.~\ref{fig:phi_br_depletions} \textbf{a} and \textbf{b}]. 
Therefore, the integrals in equation~\eqref{eq:phi_br_msr} can be restricted to $A_{1}$ and $A_{2}$:
\begin{equation}
    \dfrac{\varphi_{\text{br}}}{2\pi} = \dfrac{1}{2 l^{2}_{B}}
    \int_{A_{1}, A_{2}} r^{2} \left[ n_{|\eta_{1}|=|\eta_{2}|}(\vec{r}) -n_{\eta_{1}=\eta_{2}}(\vec{r}) \right] d\vec{r} .
    \label{eq:phi_br:integral}
\end{equation}
In these regions, the densities in Eq.~\eqref{eq:phi_br:integral} can be expressed in terms of the density depletions $d_{1\text{QH}}$ and $d_{2\text{QH}}$ caused by a single QH and two overlapping QHs~\cite{SuppMat}.
This allows us to write the braiding phase as
\begin{equation}
	\dfrac{\varphi_{\text{br}}}{2\pi} = \dfrac{1}{2 l^{2}_{B}}
	\int d\vec{\rho} \, \rho^{2} \left[ d_{2\text{QH}} (\vec{\rho}) - 2 d_{1\text{QH}} (\vec{\rho}) \right] ,
	\label{eq:phi_br_from_depletions}
\end{equation} 
in which $\vec{\rho}$ is the distance from a QH position, $d_{1\text{QH}}(\vec{\rho}) = n_{\text{b}} - n_{|\eta_{1}|=|\eta_{2}|}(\vec{\rho} + \eta_{i})$ and $d_{2\text{QH}}(\vec{\rho}) = n_{\text{b}} - n_{\eta_{1}=\eta_{2}}(\vec{\rho} + \eta_{i})$ are the aforementioned QH density depletions, w.r.t. the bulk density $n_{\text{b}} = 1/2 \pi M l^{2}_{\text{B}}$ [see Fig.~\ref{fig:phi_br_depletions} \textbf{c} and \textbf{e}].
The integration region must be large enough to ensure an appropriate decay of the density oscillations induced by the QHs. At the same time, a cutoff $\rho < R_\mathrm{max}$ is needed to avoid spurious contributions coming from the density deformations generated at the cloud boundaries.
The numerical validation of equation~\eqref{eq:phi_br_from_depletions} is shown in Fig.~\ref{fig:phi_br_depletions} \textbf{d}, \textbf{f} for the different parities $P_{N}$, and for $M=2, 1$. Residual deviations from the expected $\varphi_\text{br}$ are due to finite-size effects.

Equation~\eqref{eq:phi_br_from_depletions} constitutes an operative way to measure $\varphi_\text{br}$, which depends only on local properties in the bulk region. As such, it is robust against edge modes, which are the typical low-energy excitations due to finite-temperature effects~\cite{MacalusoCarusotto_PRA.96.043607, MacalusoCarusotto_PRA.98.013605}.
Moreover, since $d_{1\text{QH}}(\rho)$ does not depend on $P_{N}$ [see Fig.~\ref{fig:phi_br_depletions} \textbf{c} and \textbf{e}], all the information on the fusion channels is encoded in $d_{2\text{QH}}(\rho)$, which is completely different for even and odd values of $N$.
Although this dependence on $P_N$ was already known~\cite{Prodan_Haldane_PRB.80.115121, Baraban_thesis}, the key result of our work is that the depletion profiles also contain quantitative information on the braiding phase.
Note that this result holds for the QH excitations of any state in the LLL.

{\it Experimental procedure.---}
While $d_{1\text{QH}}(\rho)$ can be indifferently measured in the ground state with either a single QH or two well-separated ones~\cite{Wan_Rezayi_PRL.97.256804, Wan_Yang_PRB.77.165316}, the characterization of two overlapping QHs involves more subtleties:
First, the state in Eq.~\eqref{eq:MR_2QH} with overlapping QHs may not be the ground state in the presence of a given external potential. For instance, for odd parity $P_{N}$, having two QHs close to each other might cost more energy than just exciting a low-energy fermionic excitation at the boundary~\cite{Wen_AdvPhys.44.5.405, Milovanovic_Read_PRB.53.13559, Wan_Rezayi_PRL.97.256804}.
Furthermore, the presence of these fermionic edge modes may modify the relation between the QHs fusion channel and the particle number parity $P_{N}$ [see footnote [33] in Ref.~\cite{Wan_Yang_PRB.77.165316}].

We thus propose to proceed as follows for the measurement of $d_{2\text{QH}}(\rho)$: two QHs are created far apart, by cooling the system in the presence of pinning potentials.
The two QHs are then slowly brought closer and fused~\cite{SuppMat}. According to the general theory of topological quantum computation~\cite{Kitaev_AOP.303, Nayak_RMP.80.1083}, the fusion channel cannot change during this process, so the system is adiabatically transported into the (possibly metastable) desired state, where the depletion profile $d_{2\text{QH}}(\rho)$ is measured. Note that unless special care is taken, we can argue that in an actual experiment the QH fusion channel will be randomly chosen at each repetition~\cite{SuppMat}. Nonetheless, the non-Abelian statistics of the QHs will be still visible in the bi-peaked probability for $\varphi_{\text{br}}$. 
A rigorous proof of this statement requires numerical experiments based on a model Hamiltonian and a particular cooling mechanism; we leave it for a future study.

{\it Conclusions and Outlook.---}
In this work, we presented a scheme to assess the statistical properties of anyonic excitations which does not rely on any kind of interference.
Our protocol is based on a mathematical link between statistics and angular-momentum measurements, derived by considering rigid rotations of the anyons.
This relation further simplifies for states in the LLL, where anyonic statistics is encoded in the density profile.
Having access to the anyonic statistics without performing any interference scheme is remarkable in itself; moreover, relating statistics to density measurements makes our protocol readily applicable to state-of-the-art experiments with ultracold atoms~\cite{Cooper_etal_TopoBands_RMP} and photons~\cite{Ozawa_etal_TopoPhoto_RMP}.

Beyond the identification of the Moore-Read fusion channels, on which our scheme has been validated, the study of the two-anyons case opens several other perspectives.
For example, our method can be employed to distinguish the Moore-Read and anti-Pfaffian states, whose quasiholes have different Abelian contributions to the braiding phase~\cite{Levin_Rosenow_PRL.99.236806, Lee_Fisher_PRL.99.236807, Son_PRX.5.031027, Simon_PRB.97.121406, Feldman_PRB.98.167401, Simon_PRB.98.167402}. Moreover, it gives access to a key property in topological quantum computation~\cite{Kitaev_AOP.303, Nayak_RMP.80.1083}, namely the dependence of the braiding phase on the distance between the anyons~\cite{Baraban_Simon_numerical_analysis_MR_qh_wfs}.

Our method can also be useful for theoretical studies of states supporting anyons of unknown type.
When one can compute the matrix elements of the angular-momentum and rotation operators in the ground-state manifold, our scheme gives access to all contributions to the time-evolution operator, for any rigid rotation of the anyons.
We stress that in the case of non-Abelian anyons rigid rotations are sufficient to induce non-trivial mixing of the ground states~\cite{Nayak_RMP.80.1083}, although only a subset of the possible anyonic exchanges is accessible in this way~\cite{SuppMat}.
Therefore, we envision the possibility of a more precise theoretical characterization of the anyons, beyond the present identification of fusion channels.

Natural extensions of our analysis include other states in the LLL --like the Read-Rezayi state~\cite{Read_Rezayi_PRB.59.8084}-- or the p-wave superconductor, closely related to the Moore-Read state~\cite{ReadGreen_PRB.61.10267}.
An exciting question is whether the link between the anyonic statistics and the system density remains valid also for lattice systems~\cite{Hafezi_Lukin_PRA.76.023613, Mazza_Cirac_PRA.82.043629, Regnault_Bernevig_PRX.1.021014, Wu_PRL.113.116801, Hafezi_PRB.90.060503}; this is the subject of ongoing study~\cite{EM_etal_lattice_QHs}.

\begin{acknowledgments}
This work was supported by the EU-FET Proactive grant AQuS, Project No. 640800, and by the Autonomous Province of Trento, partially through the project ``On silicon chip quantum optics for quantum computing and secure communications" (``SiQuro"). Stimulating discussions with P. Bonderson, M. Fremling, M. O. Goerbig, V. Gurarie, C. Nayak, N. Regnault, M. Rizzi, S. H. Simon, J. K. Slingerland, and R. O. Umucal{\i}lar are warmly acknowledged.
\end{acknowledgments}

\bibliography{bibliography}

\begin{thebibliography}{60}%
\makeatletter
\providecommand \@ifxundefined [1]{%
 \@ifx{#1\undefined}
}%
\providecommand \@ifnum [1]{%
 \ifnum #1\expandafter \@firstoftwo
 \else \expandafter \@secondoftwo
 \fi
}%
\providecommand \@ifx [1]{%
 \ifx #1\expandafter \@firstoftwo
 \else \expandafter \@secondoftwo
 \fi
}%
\providecommand \natexlab [1]{#1}%
\providecommand \enquote  [1]{``#1''}%
\providecommand \bibnamefont  [1]{#1}%
\providecommand \bibfnamefont [1]{#1}%
\providecommand \citenamefont [1]{#1}%
\providecommand \href@noop [0]{\@secondoftwo}%
\providecommand \href [0]{\begingroup \@sanitize@url \@href}%
\providecommand \@href[1]{\@@startlink{#1}\@@href}%
\providecommand \@@href[1]{\endgroup#1\@@endlink}%
\providecommand \@sanitize@url [0]{\catcode `\\12\catcode `\$12\catcode
  `\&12\catcode `\#12\catcode `\^12\catcode `\_12\catcode `\%12\relax}%
\providecommand \@@startlink[1]{}%
\providecommand \@@endlink[0]{}%
\providecommand \url  [0]{\begingroup\@sanitize@url \@url }%
\providecommand \@url [1]{\endgroup\@href {#1}{\urlprefix }}%
\providecommand \urlprefix  [0]{URL }%
\providecommand \Eprint [0]{\href }%
\providecommand \doibase [0]{http://dx.doi.org/}%
\providecommand \selectlanguage [0]{\@gobble}%
\providecommand \bibinfo  [0]{\@secondoftwo}%
\providecommand \bibfield  [0]{\@secondoftwo}%
\providecommand \translation [1]{[#1]}%
\providecommand \BibitemOpen [0]{}%
\providecommand \bibitemStop [0]{}%
\providecommand \bibitemNoStop [0]{.\EOS\space}%
\providecommand \EOS [0]{\spacefactor3000\relax}%
\providecommand \BibitemShut  [1]{\csname bibitem#1\endcsname}%
\let\auto@bib@innerbib\@empty
\bibitem [{\citenamefont {Wilczek}(1982)}]{Wilczek_PRL.49.957}%
  \BibitemOpen
  \bibfield  {author} {\bibinfo {author} {\bibfnamefont {F.}~\bibnamefont
  {Wilczek}},\ }\bibfield  {title} {\enquote {\bibinfo {title} {{Quantum
  Mechanics of Fractional-Spin Particles}},}\ }\href {\doibase
  10.1103/PhysRevLett.49.957} {\bibfield  {journal} {\bibinfo  {journal} {Phys.
  Rev. Lett.}\ }\textbf {\bibinfo {volume} {49}},\ \bibinfo {pages} {957--959}
  (\bibinfo {year} {1982})}\BibitemShut {NoStop}%
\bibitem [{\citenamefont {Halperin}(1984)}]{Halperin_PRL.52.1583}%
  \BibitemOpen
  \bibfield  {author} {\bibinfo {author} {\bibfnamefont {B.~I.}\ \bibnamefont
  {Halperin}},\ }\bibfield  {title} {\enquote {\bibinfo {title} {{Statistics of
  Quasiparticles and the Hierarchy of Fractional Quantized Hall States}},}\
  }\href {\doibase 10.1103/PhysRevLett.52.1583} {\bibfield  {journal} {\bibinfo
   {journal} {Phys. Rev. Lett.}\ }\textbf {\bibinfo {volume} {52}},\ \bibinfo
  {pages} {1583--1586} (\bibinfo {year} {1984})}\BibitemShut {NoStop}%
\bibitem [{\citenamefont {Arovas}\ \emph {et~al.}(1984)\citenamefont {Arovas},
  \citenamefont {Schrieffer},\ and\ \citenamefont
  {Wilczek}}]{Arovas_Wilczek_PRL.53.722}%
  \BibitemOpen
  \bibfield  {author} {\bibinfo {author} {\bibfnamefont {D.}~\bibnamefont
  {Arovas}}, \bibinfo {author} {\bibfnamefont {J.~R.}\ \bibnamefont
  {Schrieffer}}, \ and\ \bibinfo {author} {\bibfnamefont {F.}~\bibnamefont
  {Wilczek}},\ }\bibfield  {title} {\enquote {\bibinfo {title} {{Fractional
  Statistics and the Quantum Hall Effect}},}\ }\href {\doibase
  10.1103/PhysRevLett.53.722} {\bibfield  {journal} {\bibinfo  {journal} {Phys.
  Rev. Lett.}\ }\textbf {\bibinfo {volume} {53}},\ \bibinfo {pages} {722--723}
  (\bibinfo {year} {1984})}\BibitemShut {NoStop}%
\bibitem [{\citenamefont {Leinaas}\ and\ \citenamefont
  {Myrheim}(1977)}]{Leinaas_Myrheim_Nuovo_Cimento}%
  \BibitemOpen
  \bibfield  {author} {\bibinfo {author} {\bibfnamefont {J.~M.}\ \bibnamefont
  {Leinaas}}\ and\ \bibinfo {author} {\bibfnamefont {J.}~\bibnamefont
  {Myrheim}},\ }\bibfield  {title} {\enquote {\bibinfo {title} {{On the theory
  of identical particles}},}\ }\href {\doibase 10.1007/bf02727953} {\bibfield
  {journal} {\bibinfo  {journal} {Il Nuovo Cimento B Series 11}\ }\textbf
  {\bibinfo {volume} {37}},\ \bibinfo {pages} {1--23} (\bibinfo {year}
  {1977})}\BibitemShut {NoStop}%
\bibitem [{\citenamefont {Wu}(1984)}]{Wu_PRL.52.2103}%
  \BibitemOpen
  \bibfield  {author} {\bibinfo {author} {\bibfnamefont {Y.-S.}\ \bibnamefont
  {Wu}},\ }\bibfield  {title} {\enquote {\bibinfo {title} {{General Theory for
  Quantum Statistics in Two Dimensions}},}\ }\href {\doibase
  10.1103/PhysRevLett.52.2103} {\bibfield  {journal} {\bibinfo  {journal}
  {Phys. Rev. Lett.}\ }\textbf {\bibinfo {volume} {52}},\ \bibinfo {pages}
  {2103--2106} (\bibinfo {year} {1984})}\BibitemShut {NoStop}%
\bibitem [{\citenamefont {Stern}(2008)}]{Stern_AoP}%
  \BibitemOpen
  \bibfield  {author} {\bibinfo {author} {\bibfnamefont {A.}~\bibnamefont
  {Stern}},\ }\bibfield  {title} {\enquote {\bibinfo {title} {{Anyons and the
  quantum Hall effect--A pedagogical review}},}\ }\href {\doibase
  10.1016/j.aop.2007.10.008} {\bibfield  {journal} {\bibinfo  {journal} {Annals
  of Physics}\ }\textbf {\bibinfo {volume} {323}},\ \bibinfo {pages} {204 --
  249} (\bibinfo {year} {2008})},\ \bibinfo {note} {january Special Issue
  2008}\BibitemShut {NoStop}%
\bibitem [{\citenamefont {{Tong}}(2016)}]{Tong_notes}%
  \BibitemOpen
  \bibfield  {author} {\bibinfo {author} {\bibfnamefont {D.}~\bibnamefont
  {{Tong}}},\ }\href@noop {} {\emph {\bibinfo {title} {{Lectures on the Quantum
  Hall Effect}}}}\ (\bibinfo {year} {2016})\ \Eprint
  {http://arxiv.org/abs/1606.06687} {arXiv:1606.06687} \BibitemShut {NoStop}%
\bibitem [{\citenamefont {Moore}\ and\ \citenamefont
  {Read}(1991)}]{Moore_NPB.360.2.362}%
  \BibitemOpen
  \bibfield  {author} {\bibinfo {author} {\bibfnamefont {G.}~\bibnamefont
  {Moore}}\ and\ \bibinfo {author} {\bibfnamefont {N.}~\bibnamefont {Read}},\
  }\bibfield  {title} {\enquote {\bibinfo {title} {{Nonabelions in the
  fractional quantum Hall effect}},}\ }\href {\doibase
  10.1016/0550-3213(91)90407-O} {\bibfield  {journal} {\bibinfo  {journal}
  {Nuclear Physics B}\ }\textbf {\bibinfo {volume} {360}},\ \bibinfo {pages}
  {362 -- 396} (\bibinfo {year} {1991})}\BibitemShut {NoStop}%
\bibitem [{\citenamefont {Read}\ and\ \citenamefont
  {Green}(2000)}]{ReadGreen_PRB.61.10267}%
  \BibitemOpen
  \bibfield  {author} {\bibinfo {author} {\bibfnamefont {N.}~\bibnamefont
  {Read}}\ and\ \bibinfo {author} {\bibfnamefont {D.}~\bibnamefont {Green}},\
  }\bibfield  {title} {\enquote {\bibinfo {title} {{Paired states of fermions
  in two dimensions with breaking of parity and time-reversal symmetries and
  the fractional quantum Hall effect}},}\ }\href {\doibase
  10.1103/PhysRevB.61.10267} {\bibfield  {journal} {\bibinfo  {journal} {Phys.
  Rev. B}\ }\textbf {\bibinfo {volume} {61}},\ \bibinfo {pages} {10267--10297}
  (\bibinfo {year} {2000})}\BibitemShut {NoStop}%
\bibitem [{\citenamefont {Ivanov}(2001)}]{Ivanov_PRL.86.268}%
  \BibitemOpen
  \bibfield  {author} {\bibinfo {author} {\bibfnamefont {D.~A.}\ \bibnamefont
  {Ivanov}},\ }\bibfield  {title} {\enquote {\bibinfo {title} {{Non-Abelian
  Statistics of Half-Quantum Vortices in $\mathit{p}$-Wave Superconductors}},}\
  }\href {\doibase 10.1103/PhysRevLett.86.268} {\bibfield  {journal} {\bibinfo
  {journal} {Phys. Rev. Lett.}\ }\textbf {\bibinfo {volume} {86}},\ \bibinfo
  {pages} {268--271} (\bibinfo {year} {2001})}\BibitemShut {NoStop}%
\bibitem [{\citenamefont {Read}\ and\ \citenamefont
  {Rezayi}(1999)}]{Read_Rezayi_PRB.59.8084}%
  \BibitemOpen
  \bibfield  {author} {\bibinfo {author} {\bibfnamefont {N.}~\bibnamefont
  {Read}}\ and\ \bibinfo {author} {\bibfnamefont {E.}~\bibnamefont {Rezayi}},\
  }\bibfield  {title} {\enquote {\bibinfo {title} {{Beyond paired quantum Hall
  states: Parafermions and incompressible states in the first excited Landau
  level}},}\ }\href {\doibase 10.1103/PhysRevB.59.8084} {\bibfield  {journal}
  {\bibinfo  {journal} {Phys. Rev. B}\ }\textbf {\bibinfo {volume} {59}},\
  \bibinfo {pages} {8084--8092} (\bibinfo {year} {1999})}\BibitemShut {NoStop}%
\bibitem [{\citenamefont {Nayak}\ \emph {et~al.}(2008)\citenamefont {Nayak},
  \citenamefont {Simon}, \citenamefont {Stern}, \citenamefont {Freedman},\ and\
  \citenamefont {Das~Sarma}}]{Nayak_RMP.80.1083}%
  \BibitemOpen
  \bibfield  {author} {\bibinfo {author} {\bibfnamefont {C.}~\bibnamefont
  {Nayak}}, \bibinfo {author} {\bibfnamefont {S.~H.}\ \bibnamefont {Simon}},
  \bibinfo {author} {\bibfnamefont {A.}~\bibnamefont {Stern}}, \bibinfo
  {author} {\bibfnamefont {M.}~\bibnamefont {Freedman}}, \ and\ \bibinfo
  {author} {\bibfnamefont {S.}~\bibnamefont {Das~Sarma}},\ }\bibfield  {title}
  {\enquote {\bibinfo {title} {{Non-Abelian anyons and topological quantum
  computation}},}\ }\href {\doibase 10.1103/RevModPhys.80.1083} {\bibfield
  {journal} {\bibinfo  {journal} {Rev. Mod. Phys.}\ }\textbf {\bibinfo {volume}
  {80}},\ \bibinfo {pages} {1083--1159} (\bibinfo {year} {2008})}\BibitemShut
  {NoStop}%
\bibitem [{\citenamefont {Willett}\ \emph {et~al.}(2009)\citenamefont
  {Willett}, \citenamefont {Pfeiffer},\ and\ \citenamefont
  {West}}]{Willett_PNAS.106.22}%
  \BibitemOpen
  \bibfield  {author} {\bibinfo {author} {\bibfnamefont {R.~L.}\ \bibnamefont
  {Willett}}, \bibinfo {author} {\bibfnamefont {L.~N.}\ \bibnamefont
  {Pfeiffer}}, \ and\ \bibinfo {author} {\bibfnamefont {K.~W.}\ \bibnamefont
  {West}},\ }\bibfield  {title} {\enquote {\bibinfo {title} {{Measurement of
  filling factor 5/2 quasiparticle interference with observation of charge e/4
  and e/2 period oscillations}},}\ }\href {\doibase 10.1073/pnas.0812599106}
  {\bibfield  {journal} {\bibinfo  {journal} {Proceedings of the National
  Academy of Sciences}\ }\textbf {\bibinfo {volume} {106}},\ \bibinfo {pages}
  {8853} (\bibinfo {year} {2009})}\BibitemShut {NoStop}%
\bibitem [{\citenamefont {Mourik}\ \emph {et~al.}(2012)\citenamefont {Mourik},
  \citenamefont {Zuo}, \citenamefont {Frolov}, \citenamefont {Plissard},
  \citenamefont {Bakkers},\ and\ \citenamefont
  {Kouwenhoven}}]{Mourik_Science.336.6084}%
  \BibitemOpen
  \bibfield  {author} {\bibinfo {author} {\bibfnamefont {V.}~\bibnamefont
  {Mourik}}, \bibinfo {author} {\bibfnamefont {K.}~\bibnamefont {Zuo}},
  \bibinfo {author} {\bibfnamefont {S.~M.}\ \bibnamefont {Frolov}}, \bibinfo
  {author} {\bibfnamefont {S.~R.}\ \bibnamefont {Plissard}}, \bibinfo {author}
  {\bibfnamefont {E.~P. A.~M.}\ \bibnamefont {Bakkers}}, \ and\ \bibinfo
  {author} {\bibfnamefont {L.~P.}\ \bibnamefont {Kouwenhoven}},\ }\bibfield
  {title} {\enquote {\bibinfo {title} {{Signatures of Majorana Fermions in
  Hybrid Superconductor-Semiconductor Nanowire Devices}},}\ }\href {\doibase
  10.1126/science.1222360} {\bibfield  {journal} {\bibinfo  {journal}
  {Science}\ }\textbf {\bibinfo {volume} {336}},\ \bibinfo {pages} {1003}
  (\bibinfo {year} {2012})}\BibitemShut {NoStop}%
\bibitem [{\citenamefont {Paredes}\ \emph {et~al.}(2001)\citenamefont
  {Paredes}, \citenamefont {Fedichev}, \citenamefont {Cirac},\ and\
  \citenamefont {Zoller}}]{ParedesZoller_PRL.87.010402}%
  \BibitemOpen
  \bibfield  {author} {\bibinfo {author} {\bibfnamefont {B.}~\bibnamefont
  {Paredes}}, \bibinfo {author} {\bibfnamefont {P.}~\bibnamefont {Fedichev}},
  \bibinfo {author} {\bibfnamefont {J.~I.}\ \bibnamefont {Cirac}}, \ and\
  \bibinfo {author} {\bibfnamefont {P.}~\bibnamefont {Zoller}},\ }\bibfield
  {title} {\enquote {\bibinfo {title} {{$\frac{1}{2}$-Anyons in Small Atomic
  Bose-Einstein Condensates}},}\ }\href {\doibase
  10.1103/PhysRevLett.87.010402} {\bibfield  {journal} {\bibinfo  {journal}
  {Phys. Rev. Lett.}\ }\textbf {\bibinfo {volume} {87}},\ \bibinfo {pages}
  {010402} (\bibinfo {year} {2001})}\BibitemShut {NoStop}%
\bibitem [{\citenamefont {Tserkovnyak}\ and\ \citenamefont
  {Simon}(2003)}]{Tserkovnyak_PRL.90.016802}%
  \BibitemOpen
  \bibfield  {author} {\bibinfo {author} {\bibfnamefont {Y.}~\bibnamefont
  {Tserkovnyak}}\ and\ \bibinfo {author} {\bibfnamefont {S.~H.}\ \bibnamefont
  {Simon}},\ }\bibfield  {title} {\enquote {\bibinfo {title} {{Monte Carlo
  Evaluation of Non-Abelian Statistics}},}\ }\href {\doibase
  10.1103/PhysRevLett.90.016802} {\bibfield  {journal} {\bibinfo  {journal}
  {Phys. Rev. Lett.}\ }\textbf {\bibinfo {volume} {90}},\ \bibinfo {pages}
  {016802} (\bibinfo {year} {2003})}\BibitemShut {NoStop}%
\bibitem [{\citenamefont {Baraban}\ \emph {et~al.}(2009)\citenamefont
  {Baraban}, \citenamefont {Zikos}, \citenamefont {Bonesteel},\ and\
  \citenamefont {Simon}}]{Baraban_Simon_numerical_analysis_MR_qh_wfs}%
  \BibitemOpen
  \bibfield  {author} {\bibinfo {author} {\bibfnamefont {M.}~\bibnamefont
  {Baraban}}, \bibinfo {author} {\bibfnamefont {G.}~\bibnamefont {Zikos}},
  \bibinfo {author} {\bibfnamefont {N.}~\bibnamefont {Bonesteel}}, \ and\
  \bibinfo {author} {\bibfnamefont {S.~H.}\ \bibnamefont {Simon}},\ }\bibfield
  {title} {\enquote {\bibinfo {title} {{Numerical Analysis of Quasiholes of the
  Moore-Read Wave Function}},}\ }\href {\doibase
  10.1103/PhysRevLett.103.076801} {\bibfield  {journal} {\bibinfo  {journal}
  {Phys. Rev. Lett.}\ }\textbf {\bibinfo {volume} {103}},\ \bibinfo {pages}
  {076801} (\bibinfo {year} {2009})}\BibitemShut {NoStop}%
\bibitem [{\citenamefont {Wu}\ \emph {et~al.}(2014)\citenamefont {Wu},
  \citenamefont {Estienne}, \citenamefont {Regnault},\ and\ \citenamefont
  {Bernevig}}]{Wu_PRL.113.116801}%
  \BibitemOpen
  \bibfield  {author} {\bibinfo {author} {\bibfnamefont {Y.-L.}\ \bibnamefont
  {Wu}}, \bibinfo {author} {\bibfnamefont {B.}~\bibnamefont {Estienne}},
  \bibinfo {author} {\bibfnamefont {N.}~\bibnamefont {Regnault}}, \ and\
  \bibinfo {author} {\bibfnamefont {B.~Andrei}\ \bibnamefont {Bernevig}},\
  }\bibfield  {title} {\enquote {\bibinfo {title} {{Braiding Non-Abelian
  Quasiholes in Fractional Quantum Hall States}},}\ }\href {\doibase
  10.1103/PhysRevLett.113.116801} {\bibfield  {journal} {\bibinfo  {journal}
  {Phys. Rev. Lett.}\ }\textbf {\bibinfo {volume} {113}},\ \bibinfo {pages}
  {116801} (\bibinfo {year} {2014})}\BibitemShut {NoStop}%
\bibitem [{\citenamefont {Nielsen}(2015)}]{Nielsen_PRB.91.041106}%
  \BibitemOpen
  \bibfield  {author} {\bibinfo {author} {\bibfnamefont {A.~E.~B.}\
  \bibnamefont {Nielsen}},\ }\bibfield  {title} {\enquote {\bibinfo {title}
  {{Anyon braiding in semianalytical fractional quantum Hall lattice
  models}},}\ }\href {\doibase 10.1103/PhysRevB.91.041106} {\bibfield
  {journal} {\bibinfo  {journal} {Phys. Rev. B}\ }\textbf {\bibinfo {volume}
  {91}},\ \bibinfo {pages} {041106} (\bibinfo {year} {2015})}\BibitemShut
  {NoStop}%
\bibitem [{\citenamefont {Nayak}\ and\ \citenamefont
  {Wilczek}(1996)}]{NayakWilczek_NuclPhysB479.3_1996}%
  \BibitemOpen
  \bibfield  {author} {\bibinfo {author} {\bibfnamefont {C.}~\bibnamefont
  {Nayak}}\ and\ \bibinfo {author} {\bibfnamefont {F.}~\bibnamefont
  {Wilczek}},\ }\bibfield  {title} {\enquote {\bibinfo {title} {{$2n$-quasihole
  states realize $2^{n-1}$-dimensional spinor braiding statistics in paired
  quantum Hall states}},}\ }\href {\doibase 10.1016/0550-3213(96)00430-0}
  {\bibfield  {journal} {\bibinfo  {journal} {Nuclear Physics B}\ }\textbf
  {\bibinfo {volume} {479}},\ \bibinfo {pages} {529 -- 553} (\bibinfo {year}
  {1996})}\BibitemShut {NoStop}%
\bibitem [{\citenamefont {Bonderson}\ \emph {et~al.}(2011)\citenamefont
  {Bonderson}, \citenamefont {Gurarie},\ and\ \citenamefont
  {Nayak}}]{Nayak_plasma_Ising-type_FQH}%
  \BibitemOpen
  \bibfield  {author} {\bibinfo {author} {\bibfnamefont {P.}~\bibnamefont
  {Bonderson}}, \bibinfo {author} {\bibfnamefont {V.}~\bibnamefont {Gurarie}},
  \ and\ \bibinfo {author} {\bibfnamefont {C.}~\bibnamefont {Nayak}},\
  }\bibfield  {title} {\enquote {\bibinfo {title} {{Plasma analogy and
  non-Abelian statistics for Ising-type quantum Hall states}},}\ }\href
  {\doibase 10.1103/PhysRevB.83.075303} {\bibfield  {journal} {\bibinfo
  {journal} {Phys. Rev. B}\ }\textbf {\bibinfo {volume} {83}},\ \bibinfo
  {pages} {075303} (\bibinfo {year} {2011})}\BibitemShut {NoStop}%
\bibitem [{\citenamefont {Halperin}\ \emph {et~al.}(2011)\citenamefont
  {Halperin}, \citenamefont {Stern}, \citenamefont {Neder},\ and\ \citenamefont
  {Rosenow}}]{Halperin_Rosenow_PRB.83.155440}%
  \BibitemOpen
  \bibfield  {author} {\bibinfo {author} {\bibfnamefont {B.~I.}\ \bibnamefont
  {Halperin}}, \bibinfo {author} {\bibfnamefont {A.}~\bibnamefont {Stern}},
  \bibinfo {author} {\bibfnamefont {I.}~\bibnamefont {Neder}}, \ and\ \bibinfo
  {author} {\bibfnamefont {B.}~\bibnamefont {Rosenow}},\ }\bibfield  {title}
  {\enquote {\bibinfo {title} {{Theory of the Fabry-P\'erot quantum Hall
  interferometer}},}\ }\href {\doibase 10.1103/PhysRevB.83.155440} {\bibfield
  {journal} {\bibinfo  {journal} {Phys. Rev. B}\ }\textbf {\bibinfo {volume}
  {83}},\ \bibinfo {pages} {155440} (\bibinfo {year} {2011})}\BibitemShut
  {NoStop}%
\bibitem [{\citenamefont {Campagnano}\ \emph {et~al.}(2012)\citenamefont
  {Campagnano}, \citenamefont {Zilberberg}, \citenamefont {Gornyi},
  \citenamefont {Feldman}, \citenamefont {Potter},\ and\ \citenamefont
  {Gefen}}]{Campagnano_Gefen_PRL.109.106802}%
  \BibitemOpen
  \bibfield  {author} {\bibinfo {author} {\bibfnamefont {G.}~\bibnamefont
  {Campagnano}}, \bibinfo {author} {\bibfnamefont {O.}~\bibnamefont
  {Zilberberg}}, \bibinfo {author} {\bibfnamefont {I.~V.}\ \bibnamefont
  {Gornyi}}, \bibinfo {author} {\bibfnamefont {D.~E.}\ \bibnamefont {Feldman}},
  \bibinfo {author} {\bibfnamefont {A.~C.}\ \bibnamefont {Potter}}, \ and\
  \bibinfo {author} {\bibfnamefont {Y.}~\bibnamefont {Gefen}},\ }\bibfield
  {title} {\enquote {\bibinfo {title} {{Hanbury Brown--Twiss Interference of
  Anyons}},}\ }\href {\doibase 10.1103/PhysRevLett.109.106802} {\bibfield
  {journal} {\bibinfo  {journal} {Phys. Rev. Lett.}\ }\textbf {\bibinfo
  {volume} {109}},\ \bibinfo {pages} {106802} (\bibinfo {year}
  {2012})}\BibitemShut {NoStop}%
\bibitem [{\citenamefont {Das~Sarma}\ \emph {et~al.}(2005)\citenamefont
  {Das~Sarma}, \citenamefont {Freedman},\ and\ \citenamefont
  {Nayak}}]{DasSarma_Nayak_PRL.94.166802}%
  \BibitemOpen
  \bibfield  {author} {\bibinfo {author} {\bibfnamefont {S.}~\bibnamefont
  {Das~Sarma}}, \bibinfo {author} {\bibfnamefont {M.}~\bibnamefont {Freedman}},
  \ and\ \bibinfo {author} {\bibfnamefont {C.}~\bibnamefont {Nayak}},\
  }\bibfield  {title} {\enquote {\bibinfo {title} {{Topologically Protected
  Qubits from a Possible Non-Abelian Fractional Quantum Hall State}},}\ }\href
  {\doibase 10.1103/PhysRevLett.94.166802} {\bibfield  {journal} {\bibinfo
  {journal} {Phys. Rev. Lett.}\ }\textbf {\bibinfo {volume} {94}},\ \bibinfo
  {pages} {166802} (\bibinfo {year} {2005})}\BibitemShut {NoStop}%
\bibitem [{\citenamefont {Stern}\ and\ \citenamefont
  {Halperin}(2006)}]{Stern_Halperin_PRL.96.016802}%
  \BibitemOpen
  \bibfield  {author} {\bibinfo {author} {\bibfnamefont {A.}~\bibnamefont
  {Stern}}\ and\ \bibinfo {author} {\bibfnamefont {B.~I.}\ \bibnamefont
  {Halperin}},\ }\bibfield  {title} {\enquote {\bibinfo {title} {{Proposed
  Experiments to Probe the Non-Abelian $\ensuremath{\nu}=5/2$ Quantum Hall
  State}},}\ }\href {\doibase 10.1103/PhysRevLett.96.016802} {\bibfield
  {journal} {\bibinfo  {journal} {Phys. Rev. Lett.}\ }\textbf {\bibinfo
  {volume} {96}},\ \bibinfo {pages} {016802} (\bibinfo {year}
  {2006})}\BibitemShut {NoStop}%
\bibitem [{\citenamefont {Bonderson}\ \emph {et~al.}(2006)\citenamefont
  {Bonderson}, \citenamefont {Kitaev},\ and\ \citenamefont
  {Shtengel}}]{Bonderson_Shtengel_PRL.96.016803}%
  \BibitemOpen
  \bibfield  {author} {\bibinfo {author} {\bibfnamefont {P.}~\bibnamefont
  {Bonderson}}, \bibinfo {author} {\bibfnamefont {A.}~\bibnamefont {Kitaev}}, \
  and\ \bibinfo {author} {\bibfnamefont {K.}~\bibnamefont {Shtengel}},\
  }\bibfield  {title} {\enquote {\bibinfo {title} {{Detecting Non-Abelian
  Statistics in the $\ensuremath{\nu}=5/2$ Fractional Quantum Hall State}},}\
  }\href {\doibase 10.1103/PhysRevLett.96.016803} {\bibfield  {journal}
  {\bibinfo  {journal} {Phys. Rev. Lett.}\ }\textbf {\bibinfo {volume} {96}},\
  \bibinfo {pages} {016803} (\bibinfo {year} {2006})}\BibitemShut {NoStop}%
\bibitem [{\citenamefont {Camino}\ \emph {et~al.}(2005)\citenamefont {Camino},
  \citenamefont {Zhou},\ and\ \citenamefont
  {Goldman}}]{Camino_Goldman_PRB.72.075342}%
  \BibitemOpen
  \bibfield  {author} {\bibinfo {author} {\bibfnamefont {F.~E.}\ \bibnamefont
  {Camino}}, \bibinfo {author} {\bibfnamefont {W.}~\bibnamefont {Zhou}}, \ and\
  \bibinfo {author} {\bibfnamefont {V.~J.}\ \bibnamefont {Goldman}},\
  }\bibfield  {title} {\enquote {\bibinfo {title} {{Realization of a Laughlin
  quasiparticle interferometer: Observation of fractional statistics}},}\
  }\href {\doibase 10.1103/PhysRevB.72.075342} {\bibfield  {journal} {\bibinfo
  {journal} {Phys. Rev. B}\ }\textbf {\bibinfo {volume} {72}},\ \bibinfo
  {pages} {075342} (\bibinfo {year} {2005})}\BibitemShut {NoStop}%
\bibitem [{\citenamefont {Rosenow}\ and\ \citenamefont
  {Halperin}(2007)}]{Rosenow_Halperin_PRL.98.106801}%
  \BibitemOpen
  \bibfield  {author} {\bibinfo {author} {\bibfnamefont {B.}~\bibnamefont
  {Rosenow}}\ and\ \bibinfo {author} {\bibfnamefont {B.~I.}\ \bibnamefont
  {Halperin}},\ }\bibfield  {title} {\enquote {\bibinfo {title} {{Influence of
  Interactions on Flux and Back-Gate Period of Quantum Hall
  Interferometers}},}\ }\href {\doibase 10.1103/PhysRevLett.98.106801}
  {\bibfield  {journal} {\bibinfo  {journal} {Phys. Rev. Lett.}\ }\textbf
  {\bibinfo {volume} {98}},\ \bibinfo {pages} {106801} (\bibinfo {year}
  {2007})}\BibitemShut {NoStop}%
\bibitem [{\citenamefont {Pitaevskii}\ and\ \citenamefont
  {Stringari}(2016)}]{Pitaevskii_Stringari_2016}%
  \BibitemOpen
  \bibfield  {author} {\bibinfo {author} {\bibfnamefont {L.}~\bibnamefont
  {Pitaevskii}}\ and\ \bibinfo {author} {\bibfnamefont {S.}~\bibnamefont
  {Stringari}},\ }\href {\doibase 10.1093/acprof:oso/9780198758884.001.0001}
  {\emph {\bibinfo {title} {{Bose-Einstein Condensation and Superfluidity}}}}\
  (\bibinfo  {publisher} {Oxford University Press},\ \bibinfo {year}
  {2016})\BibitemShut {NoStop}%
\bibitem [{\citenamefont {Cooper}\ \emph {et~al.}(2019)\citenamefont {Cooper},
  \citenamefont {Dalibard},\ and\ \citenamefont
  {Spielman}}]{Cooper_etal_TopoBands_RMP}%
  \BibitemOpen
  \bibfield  {author} {\bibinfo {author} {\bibfnamefont {N.~R.}\ \bibnamefont
  {Cooper}}, \bibinfo {author} {\bibfnamefont {J.}~\bibnamefont {Dalibard}}, \
  and\ \bibinfo {author} {\bibfnamefont {I.~B.}\ \bibnamefont {Spielman}},\
  }\bibfield  {title} {\enquote {\bibinfo {title} {{Topological bands for
  ultracold atoms}},}\ }\href {\doibase 10.1103/RevModPhys.91.015005}
  {\bibfield  {journal} {\bibinfo  {journal} {Rev. Mod. Phys.}\ }\textbf
  {\bibinfo {volume} {91}},\ \bibinfo {pages} {015005} (\bibinfo {year}
  {2019})}\BibitemShut {NoStop}%
\bibitem [{\citenamefont {Carusotto}\ and\ \citenamefont
  {Ciuti}(2013)}]{Carusotto_Ciuti_RMP.85.299}%
  \BibitemOpen
  \bibfield  {author} {\bibinfo {author} {\bibfnamefont {I.}~\bibnamefont
  {Carusotto}}\ and\ \bibinfo {author} {\bibfnamefont {C.}~\bibnamefont
  {Ciuti}},\ }\bibfield  {title} {\enquote {\bibinfo {title} {{Quantum fluids
  of light}},}\ }\href {\doibase 10.1103/RevModPhys.85.299} {\bibfield
  {journal} {\bibinfo  {journal} {Rev. Mod. Phys.}\ }\textbf {\bibinfo {volume}
  {85}},\ \bibinfo {pages} {299--366} (\bibinfo {year} {2013})}\BibitemShut
  {NoStop}%
\bibitem [{\citenamefont {Ozawa}\ \emph {et~al.}(2019)\citenamefont {Ozawa},
  \citenamefont {Price}, \citenamefont {Amo}, \citenamefont {Goldman},
  \citenamefont {Hafezi}, \citenamefont {Lu}, \citenamefont {Rechtsman},
  \citenamefont {Schuster}, \citenamefont {Simon}, \citenamefont {Zilberberg},\
  and\ \citenamefont {Carusotto}}]{Ozawa_etal_TopoPhoto_RMP}%
  \BibitemOpen
  \bibfield  {author} {\bibinfo {author} {\bibfnamefont {T.}~\bibnamefont
  {Ozawa}}, \bibinfo {author} {\bibfnamefont {H.~M.}\ \bibnamefont {Price}},
  \bibinfo {author} {\bibfnamefont {A.}~\bibnamefont {Amo}}, \bibinfo {author}
  {\bibfnamefont {N.}~\bibnamefont {Goldman}}, \bibinfo {author} {\bibfnamefont
  {M.}~\bibnamefont {Hafezi}}, \bibinfo {author} {\bibfnamefont
  {L.}~\bibnamefont {Lu}}, \bibinfo {author} {\bibfnamefont {M.~C.}\
  \bibnamefont {Rechtsman}}, \bibinfo {author} {\bibfnamefont {D.}~\bibnamefont
  {Schuster}}, \bibinfo {author} {\bibfnamefont {J.}~\bibnamefont {Simon}},
  \bibinfo {author} {\bibfnamefont {O.}~\bibnamefont {Zilberberg}}, \ and\
  \bibinfo {author} {\bibfnamefont {I.}~\bibnamefont {Carusotto}},\ }\bibfield
  {title} {\enquote {\bibinfo {title} {{Topological photonics}},}\ }\href
  {\doibase 10.1103/RevModPhys.91.015006} {\bibfield  {journal} {\bibinfo
  {journal} {Rev. Mod. Phys.}\ }\textbf {\bibinfo {volume} {91}},\ \bibinfo
  {pages} {015006} (\bibinfo {year} {2019})}\BibitemShut {NoStop}%
\bibitem [{\citenamefont {Macaluso}\ and\ \citenamefont
  {Carusotto}(2018)}]{MacalusoCarusotto_PRA.98.013605}%
  \BibitemOpen
  \bibfield  {author} {\bibinfo {author} {\bibfnamefont {E.}~\bibnamefont
  {Macaluso}}\ and\ \bibinfo {author} {\bibfnamefont {I.}~\bibnamefont
  {Carusotto}},\ }\bibfield  {title} {\enquote {\bibinfo {title} {{Ring-shaped
  fractional quantum Hall liquids with hard-wall potentials}},}\ }\href
  {\doibase 10.1103/PhysRevA.98.013605} {\bibfield  {journal} {\bibinfo
  {journal} {Phys. Rev. A}\ }\textbf {\bibinfo {volume} {98}},\ \bibinfo
  {pages} {013605} (\bibinfo {year} {2018})}\BibitemShut {NoStop}%
\bibitem [{\citenamefont {Wan}\ \emph {et~al.}(2006)\citenamefont {Wan},
  \citenamefont {Yang},\ and\ \citenamefont
  {Rezayi}}]{Wan_Rezayi_PRL.97.256804}%
  \BibitemOpen
  \bibfield  {author} {\bibinfo {author} {\bibfnamefont {X.}~\bibnamefont
  {Wan}}, \bibinfo {author} {\bibfnamefont {K.}~\bibnamefont {Yang}}, \ and\
  \bibinfo {author} {\bibfnamefont {E.~H.}\ \bibnamefont {Rezayi}},\ }\bibfield
   {title} {\enquote {\bibinfo {title} {{Edge Excitations and Non-Abelian
  Statistics in the Moore-Read State: A Numerical Study in the Presence of
  Coulomb Interaction and Edge Confinement}},}\ }\href {\doibase
  10.1103/PhysRevLett.97.256804} {\bibfield  {journal} {\bibinfo  {journal}
  {Phys. Rev. Lett.}\ }\textbf {\bibinfo {volume} {97}},\ \bibinfo {pages}
  {256804} (\bibinfo {year} {2006})}\BibitemShut {NoStop}%
\bibitem [{\citenamefont {Wan}\ \emph {et~al.}(2008)\citenamefont {Wan},
  \citenamefont {Hu}, \citenamefont {Rezayi},\ and\ \citenamefont
  {Yang}}]{Wan_Yang_PRB.77.165316}%
  \BibitemOpen
  \bibfield  {author} {\bibinfo {author} {\bibfnamefont {X.}~\bibnamefont
  {Wan}}, \bibinfo {author} {\bibfnamefont {Z.-X.}\ \bibnamefont {Hu}},
  \bibinfo {author} {\bibfnamefont {E.~H.}\ \bibnamefont {Rezayi}}, \ and\
  \bibinfo {author} {\bibfnamefont {K.}~\bibnamefont {Yang}},\ }\bibfield
  {title} {\enquote {\bibinfo {title} {{Fractional quantum Hall effect at
  $\ensuremath{\nu}= 5 / 2$: Ground states, non-Abelian quasiholes, and edge
  modes in a microscopic model}},}\ }\href {\doibase
  10.1103/PhysRevB.77.165316} {\bibfield  {journal} {\bibinfo  {journal} {Phys.
  Rev. B}\ }\textbf {\bibinfo {volume} {77}},\ \bibinfo {pages} {165316}
  (\bibinfo {year} {2008})}\BibitemShut {NoStop}%
\bibitem [{\citenamefont {Prodan}\ and\ \citenamefont
  {Haldane}(2009)}]{Prodan_Haldane_PRB.80.115121}%
  \BibitemOpen
  \bibfield  {author} {\bibinfo {author} {\bibfnamefont {E.}~\bibnamefont
  {Prodan}}\ and\ \bibinfo {author} {\bibfnamefont {F.~D.~M.}\ \bibnamefont
  {Haldane}},\ }\bibfield  {title} {\enquote {\bibinfo {title} {{Mapping the
  braiding properties of the Moore-Read state}},}\ }\href {\doibase
  10.1103/PhysRevB.80.115121} {\bibfield  {journal} {\bibinfo  {journal} {Phys.
  Rev. B}\ }\textbf {\bibinfo {volume} {80}},\ \bibinfo {pages} {115121}
  (\bibinfo {year} {2009})}\BibitemShut {NoStop}%
\bibitem [{Sup()}]{SuppMat}%
  \BibitemOpen
  \href@noop {} {\ }\bibinfo {note} {{See Supplemental Material.}}\BibitemShut
  {Stop}%
\bibitem [{\citenamefont {Greiter}\ \emph {et~al.}(1991)\citenamefont
  {Greiter}, \citenamefont {Wen},\ and\ \citenamefont
  {Wilczek}}]{Greiter_PRL.66.3205}%
  \BibitemOpen
  \bibfield  {author} {\bibinfo {author} {\bibfnamefont {M.}~\bibnamefont
  {Greiter}}, \bibinfo {author} {\bibfnamefont {X.-G.}\ \bibnamefont {Wen}}, \
  and\ \bibinfo {author} {\bibfnamefont {F.}~\bibnamefont {Wilczek}},\
  }\bibfield  {title} {\enquote {\bibinfo {title} {{Paired Hall state at half
  filling}},}\ }\href {\doibase 10.1103/PhysRevLett.66.3205} {\bibfield
  {journal} {\bibinfo  {journal} {Phys. Rev. Lett.}\ }\textbf {\bibinfo
  {volume} {66}},\ \bibinfo {pages} {3205--3208} (\bibinfo {year}
  {1991})}\BibitemShut {NoStop}%
\bibitem [{\citenamefont {Greiter}\ \emph {et~al.}(1992)\citenamefont
  {Greiter}, \citenamefont {Wen},\ and\ \citenamefont
  {Wilczek}}]{Greiter_NPB.374.567}%
  \BibitemOpen
  \bibfield  {author} {\bibinfo {author} {\bibfnamefont {M.}~\bibnamefont
  {Greiter}}, \bibinfo {author} {\bibfnamefont {X.-G.}\ \bibnamefont {Wen}}, \
  and\ \bibinfo {author} {\bibfnamefont {F.}~\bibnamefont {Wilczek}},\
  }\bibfield  {title} {\enquote {\bibinfo {title} {{Paired Hall states}},}\
  }\href {\doibase 10.1016/0550-3213(92)90401-V} {\bibfield  {journal}
  {\bibinfo  {journal} {Nuclear Physics B}\ }\textbf {\bibinfo {volume}
  {374}},\ \bibinfo {pages} {567 -- 614} (\bibinfo {year} {1992})}\BibitemShut
  {NoStop}%
\bibitem [{\citenamefont {Morf}(1998)}]{Morf_PRL.80.1505}%
  \BibitemOpen
  \bibfield  {author} {\bibinfo {author} {\bibfnamefont {R.~H.}\ \bibnamefont
  {Morf}},\ }\bibfield  {title} {\enquote {\bibinfo {title} {{Transition from
  Quantum Hall to Compressible States in the Second Landau Level: New Light on
  the $\ensuremath{\nu=5/2}$ Enigma}},}\ }\href {\doibase
  10.1103/PhysRevLett.80.1505} {\bibfield  {journal} {\bibinfo  {journal}
  {Phys. Rev. Lett.}\ }\textbf {\bibinfo {volume} {80}},\ \bibinfo {pages}
  {1505--1508} (\bibinfo {year} {1998})}\BibitemShut {NoStop}%
\bibitem [{\citenamefont {Rezayi}\ and\ \citenamefont
  {Haldane}(2000)}]{Rezayi_Haldane_PRL.84.4685}%
  \BibitemOpen
  \bibfield  {author} {\bibinfo {author} {\bibfnamefont {E.~H.}\ \bibnamefont
  {Rezayi}}\ and\ \bibinfo {author} {\bibfnamefont {F.~D.~M.}\ \bibnamefont
  {Haldane}},\ }\bibfield  {title} {\enquote {\bibinfo {title} {{Incompressible
  Paired Hall State, Stripe Order, and the Composite Fermion Liquid Phase in
  Half-Filled Landau Levels}},}\ }\href {\doibase 10.1103/PhysRevLett.84.4685}
  {\bibfield  {journal} {\bibinfo  {journal} {Phys. Rev. Lett.}\ }\textbf
  {\bibinfo {volume} {84}},\ \bibinfo {pages} {4685--4688} (\bibinfo {year}
  {2000})}\BibitemShut {NoStop}%
\bibitem [{\citenamefont {Ho}\ and\ \citenamefont
  {Mueller}(2002)}]{Ho_Mueller_PRL.89.050401}%
  \BibitemOpen
  \bibfield  {author} {\bibinfo {author} {\bibfnamefont {T.-L.}\ \bibnamefont
  {Ho}}\ and\ \bibinfo {author} {\bibfnamefont {E.~J.}\ \bibnamefont
  {Mueller}},\ }\bibfield  {title} {\enquote {\bibinfo {title} {{Rotating
  Spin-1 Bose Clusters}},}\ }\href {\doibase 10.1103/PhysRevLett.89.050401}
  {\bibfield  {journal} {\bibinfo  {journal} {Phys. Rev. Lett.}\ }\textbf
  {\bibinfo {volume} {89}},\ \bibinfo {pages} {050401} (\bibinfo {year}
  {2002})}\BibitemShut {NoStop}%
\bibitem [{\citenamefont {Umucal{\i}lar}\ \emph {et~al.}(2018)\citenamefont
  {Umucal{\i}lar}, \citenamefont {Macaluso}, \citenamefont {Comparin},\ and\
  \citenamefont {Carusotto}}]{Umucalilar_PRL.120.230403}%
  \BibitemOpen
  \bibfield  {author} {\bibinfo {author} {\bibfnamefont {R.~O.}\ \bibnamefont
  {Umucal{\i}lar}}, \bibinfo {author} {\bibfnamefont {E.}~\bibnamefont
  {Macaluso}}, \bibinfo {author} {\bibfnamefont {T.}~\bibnamefont {Comparin}},
  \ and\ \bibinfo {author} {\bibfnamefont {I.}~\bibnamefont {Carusotto}},\
  }\bibfield  {title} {\enquote {\bibinfo {title} {{Time-of-Flight Measurements
  as a Possible Method to Observe Anyonic Statistics}},}\ }\href {\doibase
  10.1103/PhysRevLett.120.230403} {\bibfield  {journal} {\bibinfo  {journal}
  {Phys. Rev. Lett.}\ }\textbf {\bibinfo {volume} {120}},\ \bibinfo {pages}
  {230403} (\bibinfo {year} {2018})}\BibitemShut {NoStop}%
\bibitem [{\citenamefont {Read}\ and\ \citenamefont
  {Cooper}(2003)}]{Read_Cooper_PRA.68.035601}%
  \BibitemOpen
  \bibfield  {author} {\bibinfo {author} {\bibfnamefont {N.}~\bibnamefont
  {Read}}\ and\ \bibinfo {author} {\bibfnamefont {N.~R.}\ \bibnamefont
  {Cooper}},\ }\bibfield  {title} {\enquote {\bibinfo {title} {{Free expansion
  of lowest-Landau-level states of trapped atoms: A wave-function
  microscope}},}\ }\href {\doibase 10.1103/PhysRevA.68.035601} {\bibfield
  {journal} {\bibinfo  {journal} {Phys. Rev. A}\ }\textbf {\bibinfo {volume}
  {68}},\ \bibinfo {pages} {035601} (\bibinfo {year} {2003})}\BibitemShut
  {NoStop}%
\bibitem [{\citenamefont {Macaluso}\ and\ \citenamefont
  {Carusotto}(2017)}]{MacalusoCarusotto_PRA.96.043607}%
  \BibitemOpen
  \bibfield  {author} {\bibinfo {author} {\bibfnamefont {E.}~\bibnamefont
  {Macaluso}}\ and\ \bibinfo {author} {\bibfnamefont {I.}~\bibnamefont
  {Carusotto}},\ }\bibfield  {title} {\enquote {\bibinfo {title} {{Hard-wall
  confinement of a fractional quantum Hall liquid}},}\ }\href {\doibase
  10.1103/PhysRevA.96.043607} {\bibfield  {journal} {\bibinfo  {journal} {Phys.
  Rev. A}\ }\textbf {\bibinfo {volume} {96}},\ \bibinfo {pages} {043607}
  (\bibinfo {year} {2017})}\BibitemShut {NoStop}%
\bibitem [{\citenamefont {Baraban}(2010)}]{Baraban_thesis}%
  \BibitemOpen
  \bibfield  {author} {\bibinfo {author} {\bibfnamefont {M.~S.}\ \bibnamefont
  {Baraban}},\ }\emph {\bibinfo {title} {{Low Energy Excitations in Quantum
  Condensates}}},\ \href@noop {} {Ph.D. thesis} (\bibinfo {year}
  {2010})\BibitemShut {NoStop}%
\bibitem [{\citenamefont {Wen}(1995)}]{Wen_AdvPhys.44.5.405}%
  \BibitemOpen
  \bibfield  {author} {\bibinfo {author} {\bibfnamefont {X.-G.}\ \bibnamefont
  {Wen}},\ }\bibfield  {title} {\enquote {\bibinfo {title} {{Topological orders
  and edge excitations in fractional quantum Hall states}},}\ }\href {\doibase
  10.1080/00018739500101566} {\bibfield  {journal} {\bibinfo  {journal}
  {Advances in Physics}\ }\textbf {\bibinfo {volume} {44}},\ \bibinfo {pages}
  {405--473} (\bibinfo {year} {1995})},\ \Eprint
  {http://arxiv.org/abs/https://doi.org/10.1080/00018739500101566}
  {https://doi.org/10.1080/00018739500101566} \BibitemShut {NoStop}%
\bibitem [{\citenamefont {Milovanovi\ifmmode~\acute{c}\else \'{c}\fi{}}\ and\
  \citenamefont {Read}(1996)}]{Milovanovic_Read_PRB.53.13559}%
  \BibitemOpen
  \bibfield  {author} {\bibinfo {author} {\bibfnamefont {M.}~\bibnamefont
  {Milovanovi\ifmmode~\acute{c}\else \'{c}\fi{}}}\ and\ \bibinfo {author}
  {\bibfnamefont {N.}~\bibnamefont {Read}},\ }\bibfield  {title} {\enquote
  {\bibinfo {title} {{Edge excitations of paired fractional quantum Hall
  states}},}\ }\href {\doibase 10.1103/PhysRevB.53.13559} {\bibfield  {journal}
  {\bibinfo  {journal} {Phys. Rev. B}\ }\textbf {\bibinfo {volume} {53}},\
  \bibinfo {pages} {13559--13582} (\bibinfo {year} {1996})}\BibitemShut
  {NoStop}%
\bibitem [{\citenamefont {Kitaev}(2003)}]{Kitaev_AOP.303}%
  \BibitemOpen
  \bibfield  {author} {\bibinfo {author} {\bibfnamefont {A.~Y.}\ \bibnamefont
  {Kitaev}},\ }\bibfield  {title} {\enquote {\bibinfo {title} {{Fault-tolerant
  quantum computation by anyons}},}\ }\href {\doibase
  https://doi.org/10.1016/S0003-4916(02)00018-0} {\bibfield  {journal}
  {\bibinfo  {journal} {Annals of Physics}\ }\textbf {\bibinfo {volume}
  {303}},\ \bibinfo {pages} {2 -- 30} (\bibinfo {year} {2003})}\BibitemShut
  {NoStop}%
\bibitem [{\citenamefont {Levin}\ \emph {et~al.}(2007)\citenamefont {Levin},
  \citenamefont {Halperin},\ and\ \citenamefont
  {Rosenow}}]{Levin_Rosenow_PRL.99.236806}%
  \BibitemOpen
  \bibfield  {author} {\bibinfo {author} {\bibfnamefont {M.}~\bibnamefont
  {Levin}}, \bibinfo {author} {\bibfnamefont {B.~I.}\ \bibnamefont {Halperin}},
  \ and\ \bibinfo {author} {\bibfnamefont {B.}~\bibnamefont {Rosenow}},\
  }\bibfield  {title} {\enquote {\bibinfo {title} {{Particle-Hole Symmetry and
  the Pfaffian State}},}\ }\href {\doibase 10.1103/PhysRevLett.99.236806}
  {\bibfield  {journal} {\bibinfo  {journal} {Phys. Rev. Lett.}\ }\textbf
  {\bibinfo {volume} {99}},\ \bibinfo {pages} {236806} (\bibinfo {year}
  {2007})}\BibitemShut {NoStop}%
\bibitem [{\citenamefont {Lee}\ \emph {et~al.}(2007)\citenamefont {Lee},
  \citenamefont {Ryu}, \citenamefont {Nayak},\ and\ \citenamefont
  {Fisher}}]{Lee_Fisher_PRL.99.236807}%
  \BibitemOpen
  \bibfield  {author} {\bibinfo {author} {\bibfnamefont {S.-S.}\ \bibnamefont
  {Lee}}, \bibinfo {author} {\bibfnamefont {S.}~\bibnamefont {Ryu}}, \bibinfo
  {author} {\bibfnamefont {C.}~\bibnamefont {Nayak}}, \ and\ \bibinfo {author}
  {\bibfnamefont {M.~P.~A.}\ \bibnamefont {Fisher}},\ }\bibfield  {title}
  {\enquote {\bibinfo {title} {{Particle-Hole Symmetry and the
  $\ensuremath{\nu}=\frac{5}{2}$ Quantum Hall State}},}\ }\href {\doibase
  10.1103/PhysRevLett.99.236807} {\bibfield  {journal} {\bibinfo  {journal}
  {Phys. Rev. Lett.}\ }\textbf {\bibinfo {volume} {99}},\ \bibinfo {pages}
  {236807} (\bibinfo {year} {2007})}\BibitemShut {NoStop}%
\bibitem [{\citenamefont {Son}(2015)}]{Son_PRX.5.031027}%
  \BibitemOpen
  \bibfield  {author} {\bibinfo {author} {\bibfnamefont {D.~T.}\ \bibnamefont
  {Son}},\ }\bibfield  {title} {\enquote {\bibinfo {title} {{Is the Composite
  Fermion a Dirac Particle?}}}\ }\href {\doibase 10.1103/PhysRevX.5.031027}
  {\bibfield  {journal} {\bibinfo  {journal} {Phys. Rev. X}\ }\textbf {\bibinfo
  {volume} {5}},\ \bibinfo {pages} {031027} (\bibinfo {year}
  {2015})}\BibitemShut {NoStop}%
\bibitem [{\citenamefont {Simon}(2018{\natexlab{a}})}]{Simon_PRB.97.121406}%
  \BibitemOpen
  \bibfield  {author} {\bibinfo {author} {\bibfnamefont {S.~H.}\ \bibnamefont
  {Simon}},\ }\bibfield  {title} {\enquote {\bibinfo {title} {{Interpretation
  of thermal conductance of the
  $\mathbf{\ensuremath{\nu}}=\mathbf{5}/\mathbf{2}$ edge}},}\ }\href {\doibase
  10.1103/PhysRevB.97.121406} {\bibfield  {journal} {\bibinfo  {journal} {Phys.
  Rev. B}\ }\textbf {\bibinfo {volume} {97}},\ \bibinfo {pages} {121406}
  (\bibinfo {year} {2018}{\natexlab{a}})}\BibitemShut {NoStop}%
\bibitem [{\citenamefont {Feldman}(2018)}]{Feldman_PRB.98.167401}%
  \BibitemOpen
  \bibfield  {author} {\bibinfo {author} {\bibfnamefont {D.~E.}\ \bibnamefont
  {Feldman}},\ }\bibfield  {title} {\enquote {\bibinfo {title} {{Comment on
  ``Interpretation of thermal conductance of the $\ensuremath{\nu}=5/2$
  edge''}},}\ }\href {\doibase 10.1103/PhysRevB.98.167401} {\bibfield
  {journal} {\bibinfo  {journal} {Phys. Rev. B}\ }\textbf {\bibinfo {volume}
  {98}},\ \bibinfo {pages} {167401} (\bibinfo {year} {2018})}\BibitemShut
  {NoStop}%
\bibitem [{\citenamefont {Simon}(2018{\natexlab{b}})}]{Simon_PRB.98.167402}%
  \BibitemOpen
  \bibfield  {author} {\bibinfo {author} {\bibfnamefont {S.~H.}\ \bibnamefont
  {Simon}},\ }\bibfield  {title} {\enquote {\bibinfo {title} {Reply to
  ``comment on `interpretation of thermal conductance of the
  $\ensuremath{\nu}=5/2$ edge' ''},}\ }\href {\doibase
  10.1103/PhysRevB.98.167402} {\bibfield  {journal} {\bibinfo  {journal} {Phys.
  Rev. B}\ }\textbf {\bibinfo {volume} {98}},\ \bibinfo {pages} {167402}
  (\bibinfo {year} {2018}{\natexlab{b}})}\BibitemShut {NoStop}%
\bibitem [{\citenamefont {Hafezi}\ \emph {et~al.}(2007)\citenamefont {Hafezi},
  \citenamefont {S\o{}rensen}, \citenamefont {Demler},\ and\ \citenamefont
  {Lukin}}]{Hafezi_Lukin_PRA.76.023613}%
  \BibitemOpen
  \bibfield  {author} {\bibinfo {author} {\bibfnamefont {M.}~\bibnamefont
  {Hafezi}}, \bibinfo {author} {\bibfnamefont {A.~S.}\ \bibnamefont
  {S\o{}rensen}}, \bibinfo {author} {\bibfnamefont {E.}~\bibnamefont {Demler}},
  \ and\ \bibinfo {author} {\bibfnamefont {M.~D.}\ \bibnamefont {Lukin}},\
  }\bibfield  {title} {\enquote {\bibinfo {title} {{Fractional quantum Hall
  effect in optical lattices}},}\ }\href {\doibase 10.1103/PhysRevA.76.023613}
  {\bibfield  {journal} {\bibinfo  {journal} {Phys. Rev. A}\ }\textbf {\bibinfo
  {volume} {76}},\ \bibinfo {pages} {023613} (\bibinfo {year}
  {2007})}\BibitemShut {NoStop}%
\bibitem [{\citenamefont {Mazza}\ \emph {et~al.}(2010)\citenamefont {Mazza},
  \citenamefont {Rizzi}, \citenamefont {Lewenstein},\ and\ \citenamefont
  {Cirac}}]{Mazza_Cirac_PRA.82.043629}%
  \BibitemOpen
  \bibfield  {author} {\bibinfo {author} {\bibfnamefont {L.}~\bibnamefont
  {Mazza}}, \bibinfo {author} {\bibfnamefont {M.}~\bibnamefont {Rizzi}},
  \bibinfo {author} {\bibfnamefont {M.}~\bibnamefont {Lewenstein}}, \ and\
  \bibinfo {author} {\bibfnamefont {J.~I.}\ \bibnamefont {Cirac}},\ }\bibfield
  {title} {\enquote {\bibinfo {title} {{Emerging bosons with three-body
  interactions from spin-1 atoms in optical lattices}},}\ }\href {\doibase
  10.1103/PhysRevA.82.043629} {\bibfield  {journal} {\bibinfo  {journal} {Phys.
  Rev. A}\ }\textbf {\bibinfo {volume} {82}},\ \bibinfo {pages} {043629}
  (\bibinfo {year} {2010})}\BibitemShut {NoStop}%
\bibitem [{\citenamefont {Regnault}\ and\ \citenamefont
  {Bernevig}(2011)}]{Regnault_Bernevig_PRX.1.021014}%
  \BibitemOpen
  \bibfield  {author} {\bibinfo {author} {\bibfnamefont {N.}~\bibnamefont
  {Regnault}}\ and\ \bibinfo {author} {\bibfnamefont {B.~A.}\ \bibnamefont
  {Bernevig}},\ }\bibfield  {title} {\enquote {\bibinfo {title} {{Fractional
  Chern Insulator}},}\ }\href {\doibase 10.1103/PhysRevX.1.021014} {\bibfield
  {journal} {\bibinfo  {journal} {Phys. Rev. X}\ }\textbf {\bibinfo {volume}
  {1}},\ \bibinfo {pages} {021014} (\bibinfo {year} {2011})}\BibitemShut
  {NoStop}%
\bibitem [{\citenamefont {Hafezi}\ \emph {et~al.}(2014)\citenamefont {Hafezi},
  \citenamefont {Adhikari},\ and\ \citenamefont
  {Taylor}}]{Hafezi_PRB.90.060503}%
  \BibitemOpen
  \bibfield  {author} {\bibinfo {author} {\bibfnamefont {M.}~\bibnamefont
  {Hafezi}}, \bibinfo {author} {\bibfnamefont {P.}~\bibnamefont {Adhikari}}, \
  and\ \bibinfo {author} {\bibfnamefont {J.~M.}\ \bibnamefont {Taylor}},\
  }\bibfield  {title} {\enquote {\bibinfo {title} {{Engineering three-body
  interaction and Pfaffian states in circuit QED systems}},}\ }\href {\doibase
  10.1103/PhysRevB.90.060503} {\bibfield  {journal} {\bibinfo  {journal} {Phys.
  Rev. B}\ }\textbf {\bibinfo {volume} {90}},\ \bibinfo {pages} {060503}
  (\bibinfo {year} {2014})}\BibitemShut {NoStop}%
\bibitem [{\citenamefont {Macaluso}\ \emph {et~al.}(2019)\citenamefont
  {Macaluso}, \citenamefont {Comparin}, \citenamefont {Umucal{\i}lar},
  \citenamefont {Gerster}, \citenamefont {Montangero}, \citenamefont {Rizzi},\
  and\ \citenamefont {Carusotto}}]{EM_etal_lattice_QHs}%
  \BibitemOpen
  \bibfield  {author} {\bibinfo {author} {\bibfnamefont {E.}~\bibnamefont
  {Macaluso}}, \bibinfo {author} {\bibfnamefont {T.}~\bibnamefont {Comparin}},
  \bibinfo {author} {\bibfnamefont {R.~O.}\ \bibnamefont {Umucal{\i}lar}},
  \bibinfo {author} {\bibfnamefont {M.}~\bibnamefont {Gerster}}, \bibinfo
  {author} {\bibfnamefont {S.}~\bibnamefont {Montangero}}, \bibinfo {author}
  {\bibfnamefont {M.}~\bibnamefont {Rizzi}}, \ and\ \bibinfo {author}
  {\bibfnamefont {I.}~\bibnamefont {Carusotto}},\ }\bibfield  {title} {\enquote
  {\bibinfo {title} {{Charge and statistics of lattice quasiholes from density
  measurements: a Tree Tensor Network study}},}\ }\href
  {https://arxiv.org/abs/1910.05222} {\ ,\ \bibinfo {eid} {arXiv:1910.05222}
  (\bibinfo {year} {2019})}\BibitemShut {NoStop}%
\end{thebibliography}%

\clearpage
\onecolumngrid


\section*{Supplementary material}

# Supplementary Material

## I. METHODS

### A. Rigid rotations of the anyons: ground state dynamics in the co-rotating frame

Equation (5) in the main text is derived through time-dependent perturbation theory in the reference frame $R_2$. The unperturbed Hamiltonian is $\hat{H}_1(t=0)$, while $-\theta_f \hat{L}_z/T$ is a small perturbation. We consider an initial state $|\Psi_0\rangle$ in the ground-state manifold $\mathcal{H}_{E_0}$ of $\hat{H}_1(t=0)$. The probability that the system at time $t$ has made a transition between $|\Psi_0\rangle$ and a generic state $|\psi_\beta\rangle$ reads [1]:

$$P_{\Psi_0 \to \beta} = \frac{\theta_f^2}{\hbar^2 T^2} |\langle \psi_\beta | \hat{L}_z | \Psi_0 \rangle|^2 \, \Gamma\left(\frac{E_\beta - E_0}{\hbar}, t\right), \qquad \Gamma(\omega, t) = \frac{\sin^2(\omega t/2)}{(\omega/2)^2}. \tag{S1}$$

For $T \to \infty$ and finite $t/T$, this transition probability is non-zero only for states $|\beta\rangle$ with energy $E_0$. This implies that also the time-evolved state $|\Psi_2(t)\rangle$ belongs to $\mathcal{H}_{E_0}$, and it can be decomposed on the basis $\{|\psi_\alpha\rangle\}$ as

$$|\Psi_2(t)\rangle = e^{-iE_0 t/\hbar} \sum_{\alpha=1}^m \gamma_\alpha(t) |\psi_\alpha\rangle \qquad \text{with} \qquad \gamma_\alpha(0) = \langle \psi_\alpha | \Psi_0 \rangle. \tag{S2}$$

The Schrödinger equation for $|\Psi_2\rangle$ becomes

$$i\hbar \frac{d\gamma_\alpha(t)}{dt} = -\frac{\theta_f}{T} \sum_{\beta=1}^m \mathcal{L}_{\alpha\beta} \, \gamma_\beta(t), \tag{S3}$$

with $\mathcal{L}_{\alpha\beta} = \langle \psi_\alpha | \hat{L}_z | \psi_\beta \rangle$. The solution of equation (S3) reads

$$\gamma_\alpha(t) = \sum_{\beta=1}^m \left[ e^{i\theta_f \mathcal{L} t/(\hbar T)} \right]_{\alpha,\beta} \gamma_\beta(0). \tag{S4}$$

This justifies equations (5) and (6) in the main text, for the time-evolved states $|\Psi_2(T)\rangle$ and $|\Psi_1(T)\rangle$.

### B. Analytic wave functions for two Moore-Read quasiholes

For the Moore-Read (MR) 2QH states we consider the following wave functions:

$$\Psi^{\text{2QH}}(\eta; z) = \text{Pf}(W) \prod_{i<j}(z_i - z_j)^M e^{-\frac{1}{4}\sum_i^N |z_i|^2}, \tag{S5}$$

where the form of the anti-symmetric matrix $W$ depends on the parity $P_N$ of the particle number $N$ [2]. If $P_N = 0$ ($N$ even), $W$ is a $N \times N$ matrix with

$$W_{ij} = \begin{cases} 0 & i = j \\ \frac{(\eta_1 - z_i)(\eta_2 - z_j) + (i \leftrightarrow j)}{z_i - z_j} & i \neq j \end{cases} \tag{S6}$$

while if $P_N = 1$ ($N$ odd), $W$ is a $(N+1) \times (N+1)$ matrix with

$$W_{ij} = \begin{cases} 0 & i = j \\ \frac{(\eta_1 - z_i)(\eta_2 - z_j) + (i \leftrightarrow j)}{z_i - z_j} & i, j < N+1 \\ 1 & i = N+1, j \leq N \\ -1 & j = N+1, i \leq N \end{cases} \tag{S7}$$

These wave functions strongly depend on the parity $P_N$. For $P_N = 0$, $\text{Pf}(W)$ corresponds to a single Pfaffian depending on all the $N$ particle coordinates. For $P_N = 1$, $\text{Pf}(W)$ is the sum (with appropriate signs) of $N$ different Pfaffians, where each Pfaffian depends only on $N-1$ particle coordinates. Due to this distinction, the mean angular momenta of the states with $\eta_1 = \eta_2 = 0$ depends on the parity $P_N$:

$$\langle \hat{L}_z \rangle_{0,0} = \hbar \left[ \frac{MN(N-1)}{2} + \frac{N}{2} - \frac{P_N}{2} \right]. \tag{S8}$$

This distinction is crucial in our study of the fusion channels characterizing the non-Abelian QHs of the MR state.

### C. Metropolis Monte Carlo method

Results reported in Table I and Fig. 2 of the main text required the evaluation of the expectation values of some spatial dependent observables $\mathcal{O}(\vec{r})$ on different states. We recast these expectation values as averages over the probability distribution given by the squared modulus of the wave functions of interest,

$$\frac{\langle \Psi | \mathcal{O}(\vec{r}) | \Psi \rangle}{\langle \Psi | \Psi \rangle} = \frac{\int \mathcal{O}(\vec{r}) \, |\Psi(z_1, \ldots, z_N)|^2 \, dz_1 \ldots dz_N}{\int |\Psi(z_1, \ldots, z_N)|^2 \, dz_1 \ldots dz_N}, \tag{S9}$$

and we compute these integrals through the Metropolis Monte Carlo technique [3, 4], leading to results which are exact within statistical errors. This is an established method also in the fractional quantum Hall context, see for instance Refs. [5–7]. Note that we treat the Pfaffian factors in $|\Psi|^2$ directly, without making use of their integral representation introduced in Ref. [8].

## II. RIGID ROTATIONS OF THE ANYONS: ALTERNATIVE DERIVATION

While in the main text we derive the effect of rigid rotations of the anyons by solving the dynamics of the system in the co-rotating reference frame, here we show that the same results can be obtained through the application of the adiabatic theorem in a suitable gauge.

We consider the Hamiltonian $\hat{H}(t) = \hat{H}(\lambda_1(t), \ldots \lambda_n(t))$, where $\vec{\lambda}(t)$ is a set of time-dependent parameters. We denote by $\mathcal{H}_{E_0(t)}$ the (possibly degenerate) subspace of ground states of $\hat{H}(t)$, spanned by the basis $\{|\psi_\alpha(t)\rangle\}$, and we assume that an energy gap separates the ground-state energy $E_0(t)$ from the other eigenstates at any time $t$. Then, according to the adiabatic theorem [9, 10], if we start at $t = 0$ in the state $|\Psi(0)\rangle \in \mathcal{H}_{E_0(0)}$, the time-evolved state $|\Psi(T)\rangle$ will be given by

$$|\Psi(T)\rangle = \mathcal{U}(T)|\Psi(0)\rangle \in \mathcal{H}_{E_0(T)}. \tag{S10}$$

The unitary transformation $\mathcal{U}(T)$ is gauge invariant if and only if the adiabatic process describes a closed path in parameter space –i.e. $\hat{H}(T) = \hat{H}(0)$– and it is in general the product of three contributions: (i) The dynamical phase factor

$$\mathcal{U}_{\text{dyn}} = \exp\left[ -\frac{i}{\hbar} \int_0^T E_0(t) dt \right]. \tag{S11}$$

(ii) The geometric contribution

$$\mathcal{U}_{\text{B}} = \mathcal{P} \exp\left[ i \int_0^T \mathcal{A}(t) dt \right], \tag{S12}$$

known as Berry matrix [10, 11], in which $\mathcal{P}$ stands for path –or time– ordering and the Berry connection $\mathcal{A}(t)$ has matrix elements

$$\mathcal{A}_{\alpha\beta}(t) = i \langle \psi_\alpha(t) | \frac{d}{dt} | \psi_\beta(t) \rangle = \mathcal{A}_{\alpha\beta}^{\vec{\lambda}} \cdot \frac{d\vec{\lambda}(t)}{dt},$$
$$\mathcal{A}_{\alpha\beta}^{\vec{\lambda}} = i \langle \psi_\alpha(\vec{\lambda}) | \frac{\partial}{\partial \vec{\lambda}} | \psi_\beta(\vec{\lambda}) \rangle. \tag{S13}$$

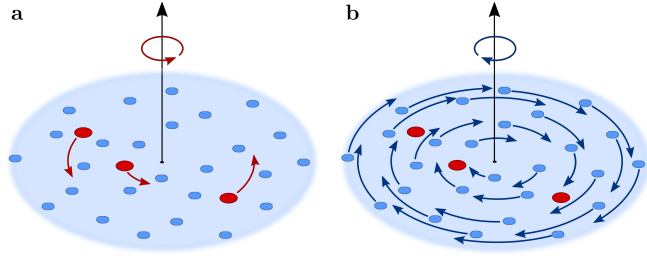

FIG. S1. Schematic pictures of the rigid rotation of the anyons (**a**) and of the particles (**b**).

(iii) A unitary transformation $\mathcal{B}$ taking into account the change of the basis states during the adiabatic process, with matrix elements

$$\mathcal{B}_{\alpha\beta} = \langle\psi_\alpha(0)|\psi_\beta(T)\rangle, \tag{S14}$$

so that $|\psi_\alpha(T)\rangle = \mathcal{B}|\psi_\alpha(0)\rangle$.

When considering the adiabatic braiding/exchange of anyons, the parameters $\vec{\lambda}(t)$ are generally identified with positions of some external potentials pinning the anyons at coordinates $\eta_\mu(t)$. For the special case of rigid rotations of the anyons, where $\eta_\mu(t) = \eta_\mu(0)e^{i\theta(t)}$ [Fig. S1 **a**], the only relevant parameter is the rotation angle $\theta(t) = \theta_f t/T$.

Moreover, we assume that

$$\hat{H}(z_j, t) = \hat{\mathcal{R}}^\dagger_{-\theta(t)} \hat{H}(z_j, 0) \hat{\mathcal{R}}_{-\theta(t)} = \hat{H}(z_j e^{-i\theta(t)}, 0) \tag{S15}$$

at any time $t$, where $\hat{\mathcal{R}}_{\theta(t)} = \exp\left[-i\theta(t)\hat{L}_z/\hbar\right]$ is the particle rotation operator. Equation (S15) is generally satisfied by FQH Hamiltonians with potential terms which only depends on the distance $|z_j - \eta_\mu(t)|$ of the particles from the anyonic coordinates, as those considered in the main text. Through equation (S15), the non-dynamical contributions to $\mathcal{U}(T)$ for rigid rotations can be rewritten in a simpler fashion. We define the basis states at time $t$ starting from those of the initial basis and by rotating all the particles [Fig. S1 **b**], i.e.

$$|\psi_\alpha(t)\rangle = \hat{\mathcal{R}}_{\theta(t)}|\psi_\alpha(0)\rangle. \tag{S16}$$

This particle-rotation (PR) basis differs from the one obtained by following the time evolution of the anyon coordinates, i.e. by substituting $\eta_\mu(0)$ with $\eta_\mu(t)$ in the states of the basis at time 0 –e.g. in [8]. With this second approach one would obtain the anyon-rotation (AR) basis

$$|\tilde{\psi}_\alpha(t)\rangle = \mathcal{S}_{\theta(t)}|\tilde{\psi}_\alpha(0)\rangle, \tag{S17}$$

where $\mathcal{S}_{\theta(t)}$ is the anyon-rotation transformation that maps a generic wave function $\Psi(\eta; z)$ into $\Psi(\eta e^{i\theta(t)}; z)$. Although in general $|\psi_\alpha(t)\rangle \neq |\tilde{\psi}_\alpha(t)\rangle$, equation (S15) ensures that the two basis span the same subspace $\mathcal{H}_{E_0(t)}$ at any time $t$.

In the PR basis, one may express the Berry connection (S13) in terms of the angular momentum $\hat{L}_z$,

$$\mathcal{A}^\theta_{\alpha\beta} = \frac{1}{\hbar}\langle\psi_\alpha(\theta)|\hat{L}_z|\psi_\beta(\theta)\rangle = \frac{1}{\hbar}\langle\psi_\alpha(0)|\hat{\mathcal{R}}^\dagger_{\theta(t)}\hat{L}_z\hat{\mathcal{R}}_{\theta(t)}|\psi_\beta(0)\rangle = \frac{1}{\hbar}\langle\psi_\alpha(0)|\hat{L}_z|\psi_\beta(0)\rangle, \tag{S18}$$

where $\theta \in [0, \theta_f]$, and where we used the fact that $\hat{L}_z$ and $\hat{\mathcal{R}}_{\theta(t)}$ commute. Therefore the Berry matrix reads

$$\mathcal{U}_\text{B} = e^{i\theta_f \mathcal{L}/\hbar}, \tag{S19}$$

where the matrix elements of $\mathcal{L}$ are the angular momentum matrix elements evaluated on the states of the basis at initial time (as in the main text), $\mathcal{L}_{\alpha\beta} = \langle\psi_\alpha(0)|\hat{L}_z|\psi_\beta(0)\rangle$.

Note that here the integration appearing in equation (S12) only gives an overall constant factor $\theta_f$. Therefore, by working in the PR basis, one can obtain the Berry matrix $\mathcal{U}_\text{B}$ through the measurement of the $\hat{L}_z$ matrix elements at $\theta = 0$, i.e. without performing any actual rotation.

Moreover, the basis-change matrix $\mathcal{B}$ has elements

$$\mathcal{B}_{\alpha\beta} = \langle\psi_\alpha(0)|\psi_\beta(T)\rangle = \langle\psi_\alpha(0)|\hat{\mathcal{R}}_{\theta_f}|\psi_\beta(0)\rangle = \langle\psi_\alpha(0)|e^{-i\theta_f \hat{L}_z/\hbar}|\psi_\beta(0)\rangle, \tag{S20}$$

and this imposes the constraint $\mathcal{B}(2\pi k) \equiv \mathbb{1}$, where $\mathbb{1}$ denotes the identity operator and $k$ is an integer number. This constraint comes from the fact that the basis states at different angles are obtained by acting only on the particle coordinates and that any admissible wave function must be single-valued with respect to them. Therefore, when considering $2\pi k$-rotations in the PR basis, all non-dynamical contributions to $\mathcal{U}(T)$ –including those coming from the anyonic braiding statistics– must be in the Berry matrix $\mathcal{U}_B$. Note that this is true if and only if $\theta_f = 2\pi k$. On the contrary, wave functions can be in general multi-valued in the anyon coordinates $\eta_\mu$. Therefore, if one chooses the AR basis, the unitary transformation $\mathcal{U}(T)$ can in principle take non-trivial contributions from basis-change matrix $\tilde{\mathcal{B}} = \langle \tilde{\psi}_\alpha(0) | \tilde{\psi}_\beta(T) \rangle = \langle \tilde{\psi}_\alpha(0) | \mathcal{S}_{\theta_f} | \tilde{\psi}_\beta(0) \rangle$ also in the case of $2\pi k$-rotations.

A final remark concerns the case of anyonic configurations in which the set of coordinates $\eta_\mu$ maps into itself for rotation angles $\theta_f \neq 2\pi k$ [see Fig. 1 **c** in the main text]. In this case, the basis-change matrix $\mathcal{B}$ is not proportional to the identity, in general, and it can also have non-vanishing off-diagonal matrix elements. From the experimental point of view this is not the optimal situation, since one cannot easily measure the entries of $\mathcal{B}$. However, rigid rotations of angles $\theta_f \neq 2\pi k$ can be extremely useful in view of theoretical applications of our study. In particular, by computing $\mathcal{U}_B$ and $\mathcal{B}$ for these rotations, one can demonstrate that the unitary transformation $\mathcal{U}(T)$ has non-vanishing off-diagonal matrix elements, without performing any kind of time evolution.

We stress that the expressions for $\mathcal{U}_B$ and $\mathcal{B}$ in the PR basis [see equations (S19) and (S20)] are exactly the same as those obtained in the main text by solving the ground state dynamics in the reference frame co-rotating with the anyons.

## III. AHARONOV-BOHM PHASE DUE TO PARTICLE ROTATION

Here we explicitly compute the non-topological (NT) contribution to the Berry phase $\varphi_B = \theta_f \mathcal{L}/\hbar$ acquired by the MR 2QH state after a $2\pi$-rotation of the anyons. Since in the PR basis $\varphi_B$ can be expressed in terms of the angular momentum expectation value taken on the 2QH state, we can write

$$\varphi_B(\eta_1, \eta_2) = \frac{2\pi}{\hbar} \langle \hat{L}_z \rangle_{\eta_1, \eta_2} = \varphi_{NT}(\eta_1, \eta_2) + \varphi_{br}, \tag{S21}$$

in which $\varphi_{NT}(\eta_1, \eta_2)$ is the aforementioned non-topological contribution to $\varphi_B$ and $\eta_1$, $\eta_2$ denote the QH coordinates.

To find an analytic expression for $\varphi_{NT}(\eta_1, \eta_2)$, we consider the generic state with far apart QHs at positions $\eta_1$, $\eta_2$ [with $|\eta_1| < |\eta_2|$], for a large number of particles $N \gg 1$. This approximation allows us assume that: i) The density $n(\mathbf{r})$ of the FQH cloud is a constant $n_b = 1/2\pi M l_B^2$ in the interval $0 \leq |\mathbf{r}| \leq R_{cl}$, where $R_{cl} = R_{cl}(N)$ denotes the semi-classical radius of the cloud, and it is equal to $0$ elsewhere. In other words, we are neglecting the density bump at the cloud boundary. ii) The QH size is negligible, and therefore each QH can be seen as an additional magnetic flux $\Phi_{QH} = \pi\hbar/e = \Phi_0/2$, where $\Phi_0 = 2\pi\hbar/e$ is the quantum flux, located at the QH position. As a consequence of ii), the QH positions $\eta_1$ and $\eta_2$ divide the FQH cloud in three regions: A circular region $S_1$ of radius $|\eta_1|$, a ring-shaped region $S_2$ with inner radius $|\eta_1|$ and outer radius $|\eta_2|$, and a ring-shaped region $S_3$ extending from $|\eta_2|$ to $R_{cl}$ [see Fig. S2].

For adiabatic processes in which charged objects move in presence of magnetic fields, the non-topological phase $\varphi_{NT}$ acquired by the many-body wave function is an Aharonov-Bohm (AB) phase [12]. Therefore equation (S21) becomes

$$\frac{2\pi}{\hbar} \langle \hat{L}_z \rangle_{\eta_1, \eta_2} = \varphi_{AB}(\eta_1, \eta_2) + \varphi_{br}. \tag{S22}$$

where $\varphi_{AB}(\eta_1, \eta_2)$ is not the AB phase associated with the rotation of the QHs, but rather the one due to the rotation of the particles forming the MR state. This is a consequence of working in the PR basis, in which anyons are fixed and particles move [see Fig. S1].

For the simplest case of a charged particle moving along a closed path in presence of a constant magnetic field, the AB phase is $q\Phi/\hbar$, where $q$ and $\Phi$ denote the particle charge and the magnetic flux passing through the surface enclosed by the particle path, respectively. Here we consider a charge density distributed over the FQH cloud, rather than individual particles. Under our previous assumption that the density is constant for all $r \leq R_{cl}$, the AB phase $\varphi_{AB}(\eta_1, \eta_2)$ is $\int_0^{R_{cl}} [2\pi r \, q n_b \Phi(r)/\hbar] \, dr$, where $q n_b$ is the charge density and where the flux $\Phi(r)$ seen at radius $r$ depends on the region $S_i$ to which $r$ belongs. Particles in $S_1$ only feel the flux $\Phi_{S_1}(r) = \pi r^2 \rho_B \Phi_0$ due to the external magnetic field, where $\pi r^2$ is the surface enclosed by the ring and $\rho_B = 1/2\pi l_B^2$ is the flux density. Particles in $S_2$ feel the external-field flux plus an additional magnetic flux $\Phi_{QH}$ due to the fact they encircle a QH. Particles in $S_3$ encircle two QHs, and thus they feel two additional fluxes: $\Phi_{S_3}(r) = \pi r^2 \rho_B \Phi_0 + 2\Phi_{QH}$. The AB phases associated with the rotation of the charge density in the three different regions are

$$\varphi_{AB}^{S_1}(\eta_1) = \frac{q n_b}{\hbar} \int_0^{|\eta_1|} dr \, 2\pi r \, \Phi_{S_1}(r) = \frac{q n_b 2\pi^2 B}{\hbar} \frac{|\eta_1|^4}{4}, \tag{S23}$$

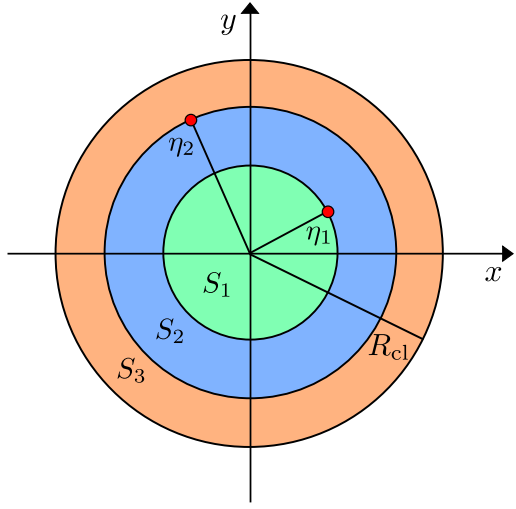

FIG. S2. Schematic picture of the regions $S_1$, $S_2$ and $S_3$ of the FQH cloud defined by the quasihole positions $\eta_1$, $\eta_2$ (red dots) in the large $N$ limit.

$$\varphi_{\text{AB}}^{S_2}(\eta_1, \eta_2) = \frac{qn_{\text{b}}}{\hbar} \int_{|\eta_1|}^{|\eta_2|} dr\, 2\pi r \Phi_{S_2}(r) = \frac{qn_{\text{b}} 2\pi^2 B}{\hbar}\left(\frac{|\eta_2|^4}{4} - \frac{|\eta_1|^4}{4}\right) + \frac{qn_{\text{b}}\pi\Phi_0}{\hbar}\left(\frac{|\eta_2|^2}{2} - \frac{|\eta_1|^2}{2}\right), \quad (\text{S24})$$

$$\varphi_{\text{AB}}^{S_3}(\eta_2) = \frac{qn_{\text{b}}}{\hbar} \int_{|\eta_2|}^{R_{\text{cl}}} dr\, 2\pi r \Phi_{S_3}(r) = \frac{qn_{\text{b}} 2\pi^2 B}{\hbar}\left(\frac{R_{\text{cl}}^4}{4} - \frac{|\eta_2|^4}{4}\right) + \frac{qn_{\text{b}} 2\pi\Phi_0}{\hbar}\left(\frac{R_{\text{cl}}^2}{2} - \frac{|\eta_2|^2}{2}\right), \quad (\text{S25})$$

and the global AB phase reads

$$\begin{aligned}\varphi_{\text{AB}}(\eta_1,\eta_2) &= \varphi_{\text{AB}}^{S_1}(\eta_1) + \varphi_{\text{AB}}^{S_2}(\eta_1,\eta_2) + \varphi_{\text{AB}}^{S_3}(\eta_2) \\ &= \frac{qn_{\text{b}} 2\pi^2 B}{\hbar c}\frac{R_{\text{cl}}^4}{4} + \frac{qn_{\text{b}} 2\pi\Phi_0}{\hbar c}\frac{R_{\text{cl}}^2}{2} - \frac{qn_{\text{b}}\pi\Phi_0}{\hbar c}\left(\frac{|\eta_1|^2}{2} + \frac{|\eta_2|^2}{2}\right) \\ &= K(M,N) - 2\pi\frac{1}{2M}\left(\frac{|\eta_1|^2}{2l_B^2} + \frac{|\eta_2|^2}{2l_B^2}\right),\end{aligned} \quad (\text{S26})$$

where in the last equality we used $n_{\text{b}} = 1/2\pi M l_B^2$, $\Phi_0 = 2\pi l_B^2 B$ and $q = e$. Here $K(M, N)$ is a constant which does not depend on $\eta_1$ and $\eta_2$. The second term on the right-hand side of equation (S26) is the AB phase associated with the rotation of the two QHs [see for instance Ref. [8]], which means that the dependence of $\langle \hat{L}_z \rangle_{\eta_1, \eta_2}$ on the QH positions allows one to compute the fractional charge of MR QHs. This is also an indication of the correctness of our reasoning: particle and QH rotations are indeed related.

The prediction for $K(M, N)$, however, is not reliable. This is easily understood by observing that in the large-$N$ approximation the semi-classical radius of the FQH cloud $R_{\text{cl}}$ does not depend on the number $N_{\text{QH}}$ of QH excitations. If our expression for $K(M, N)$ were correct, then also the mean angular momentum of the MR state with $\eta_1 = \eta_2 = 0$ would not depend on $N_{\text{QH}}$, but we know that this is not the case.

The issue is solved by rewriting equation (S22) for the case of $\eta_1 = \eta_2 = 0$, since in this case the particle rotation does not induce any exchange of the QHs and so $\varphi_{\text{br}} = 0$. We find

$$\frac{2\pi}{\hbar}\langle \hat{L}_z \rangle_{0,0} = \varphi_{\text{AB}}(0,0) + \varphi_{\text{br}} = \varphi_{\text{AB}}(0,0), \quad (\text{S27})$$

which implies that $K(M, N)$ must be computed as the mean angular momentum of the MR state having both QHs located in the origin: $K(M, N) = (2\pi/\hbar)\langle \hat{L}_z \rangle_{0,0}$. Equation (S22) then becomes

$$\frac{2\pi}{\hbar}\langle \hat{L}_z \rangle_{\eta_1,\eta_2} = \frac{2\pi}{\hbar}\langle \hat{L}_z \rangle_{0,0} - 2\pi\frac{1}{2M}\left(\frac{|\eta_1|^2}{2l_B^2} + \frac{|\eta_2|^2}{2l_B^2}\right) + \varphi_{\text{br}}, \quad (\text{S28})$$

and the braiding phase of MR non-Abelian QHs is related to the angular momentum via

$$\varphi_{\text{br}} = \frac{2\pi}{\hbar}\langle \hat{L}_z \rangle_{\eta_1,\eta_2} - \frac{2\pi}{\hbar}\langle \hat{L}_z \rangle_{0,0} + 2\pi\frac{1}{2M}\left(\frac{|\eta_1|^2}{2l_B^2} + \frac{|\eta_2|^2}{2l_B^2}\right). \quad (\text{S29})$$

Finally, we notice that if we consider a state with QH at positions $\eta_1$, $\eta_2$ such that $|\eta_1| = |\eta_2|$, the last two terms on the right-hand side of equation (S29) correspond to the mean angular momentum of the state with QHs on top of each other at distance $|\eta|$ from the center of the cloud. So we have

$$\varphi_{\text{br}} = \frac{2\pi}{\hbar} \langle \hat{L}_z \rangle_{|\eta_1|=|\eta_2|} - \frac{2\pi}{\hbar} \langle \hat{L}_z \rangle_{\eta_1=\eta_2}, \tag{S30}$$

which corresponds to equation (13) of the main text.

The calculations of this section are easily generalized to the QHs of the Laughlin state, with the only difference that the corresponding QH magnetic flux is $\Phi_{\text{QH}} = \Phi_0$ instead of $\Phi_0/2$.

## IV. RELATION BETWEEN QUASIHOLE BRAIDING PHASE AND DENSITY DEPLETIONS

In this Section we show how to express the QH braiding phase in terms of the QH density depletions (equation (13) in the main text), starting from its expression in term of the mean square radii (equation (11) in the main text).

First we rewrite equation (11) in the main text as

$$\begin{aligned}
\frac{\varphi_{\text{br}}}{2\pi} &= \frac{N}{2l_B^2} \left[ \langle r^2 \rangle_{|\eta_1|=|\eta_2|} - \langle r^2 \rangle_{\eta_1=\eta_2} \right] \\
&= \frac{N}{2l_B^2} \frac{1}{N} \int_{\mathbb{R}^2} r^2 \left[ n_{|\eta_1|=|\eta_2|}(\vec{r}) - n_{\eta_1=\eta_2}(\vec{r}) \right] d\vec{r} \\
&= \frac{1}{2l_B^2} \int_{A_1, A_2} r^2 \left[ n_{|\eta_1|=|\eta_2|}(\vec{r}) - n_{\eta_1=\eta_2}(\vec{r}) \right] d\vec{r},
\end{aligned} \tag{S31}$$

where in the last equivalence we used the fact that the densities $n_{|\eta_1|=|\eta_2|}(\vec{r})$ and $n_{\eta_1=\eta_2}(\vec{r})$ are equal over the whole 2D plane, except for the surfaces $A_1$ and $A_2$ surrounding the QHs [red circles in Fig. 2 a and b in the main text]. Densities $n_{|\eta_1|=|\eta_2|}(\vec{r})$ and $n_{\eta_1=\eta_2}(\vec{r})$ in the regions $A_1$ and $A_2$ can be rewritten as

$$\begin{aligned}
n_{|\eta_1|=|\eta_2|}(\vec{r}) &= n_{\text{b}} - d_{1\text{QH}}(\vec{r} - \eta_j) \quad \forall \vec{r} \in A_j,\ j=1,2 \\
n_{\eta_1=\eta_2}(\vec{r}) &= n_{\text{b}} - d_{2\text{QH}}(\vec{r} - \eta_1) \quad \forall \vec{r} \in A_1, \\
n_{\eta_1=\eta_2}(\vec{r}) &= n_{\text{b}} \qquad\qquad\qquad\qquad \forall \vec{r} \in A_2,
\end{aligned} \tag{S32}$$

where $n_{\text{b}} = 1/2\pi M l_B^2$ is the bulk density, and $d_{1\text{QH}}(\vec{r})$ and $d_{2\text{QH}}(\vec{r})$ represent the density depletions caused by a single QH or by two QHs on top of each other, respectively. By inserting these expressions in the previous integral one obtains

$$\begin{aligned}
\frac{\varphi_{\text{br}}}{2\pi} &= \frac{1}{2l_B^2} \left\{ \int_{A_1} r^2 \left[ d_{2\text{QH}}(\vec{r} - \eta_1) - d_{1\text{QH}}(\vec{r} - \eta_1) \right] d\vec{r} - \int_{A_2} r^2 d_{1\text{QH}}(\vec{r} - \eta_2) d\vec{r} \right\} = \\
&= \frac{1}{2l_B^2} \int_{A_1} r^2 \left[ d_{2\text{QH}}(\vec{r} - \eta_1) - 2d_{1\text{QH}}(\vec{r} - \eta_1) \right] d\vec{r},
\end{aligned} \tag{S33}$$

where we used the facts that $|\eta_1| = |\eta_2|$ and that the integrals do not depend on the angle distinguishing $A_1$ and $A_2$. By changing variables to $\vec{\rho} \equiv \vec{r} - \eta_1$, we obtain

$$\begin{aligned}
\frac{\varphi_{\text{br}}}{2\pi} &= \frac{1}{2l_B^2} \int_{A_1} (\vec{\rho} - \eta_1)^2 \left[ d_{2\text{QH}}(\vec{\rho}) - 2d_{1\text{QH}}(\vec{\rho}) \right] d\vec{\rho} \\
&= \frac{1}{2l_B^2} \int_0^{2\pi} d\theta \int_0^{R_{\max}} d\rho\, \rho\, (\rho^2 - 2\rho|\eta_1|\cos\theta + |\eta_1|^2) \left[ d_{2\text{QH}}(\rho) - 2d_{1\text{QH}}(\rho) \right],
\end{aligned} \tag{S34}$$

where $R_{\max}$ is the radius of the disk-shaped surface $A_1$. This radius must be small enough to avoid the overlap between the different $A_i$ regions and the spurious contributions from the density deformations at the FQH cloud boundary, but large enough to ensure an appropriate damping of the density oscillations induced by the QHs on top of the bulk density.

Finally, we note that the integrate density depletion for two QHs on top of each other and for two QHs far away is the same, i.e.

$$\int_0^{2\pi} d\theta \int_0^{R_{\max}} d\rho\, \rho\, \left[ d_{2\text{QH}}(\rho) - 2d_{1\text{QH}}(\rho) \right] = 0, \tag{S35}$$

and that the integral of $\cos\theta$ vanishes in the interval $[0, 2\pi]$. Therefore, the previous expression for the braiding phase simplifies and corresponds to equation (13) in the main text.

The reasoning in this section applies to all QH excitations of FQH states in the lowest Landau level (LLL), since it only requires the validity of the expression relating the mean angular momentum of a state in the LLL to its mean square radius [7].

## V. FUSION OF THE MOORE-READ QUASIHOLES

In the main text we mentioned that a qualitative proof of the existence of multiple fusion channels can be obtained by bringing the MR QHs close to each other and by looking at their density profiles. Here we provide numerical results, based on Monte Carlo calculations, showing how the density profile of the MR QHs changes as a function of the distance between them [see Fig. S3].

The density depletions created by QHs far apart from each other do not depend on the parity $P_N$ of the particle number, as expected [see Fig. S3 **a**]. In this respect, is it interesting to notice that the density profiles seem to be parity-independent as long as the two QHs do not significantly overlap [see Fig. S3 **b**]. If then one brings the QHs closer to each other, they start to hybridize and the resulting density profile strongly depends on $P_N$ [see Fig. S3 **c**]. Finally, the difference between the $P_N = 0, 1$ density profiles reaches is maximum when the QHs are on top of each other [see Fig. S3 **d**]. For $P_N = 0$ one has vanishing density at the QHs position, while for $P_N = 1$ the minimum value of the density is approximately one half of the bulk density $n_b$ and it is not located at the QH positions.

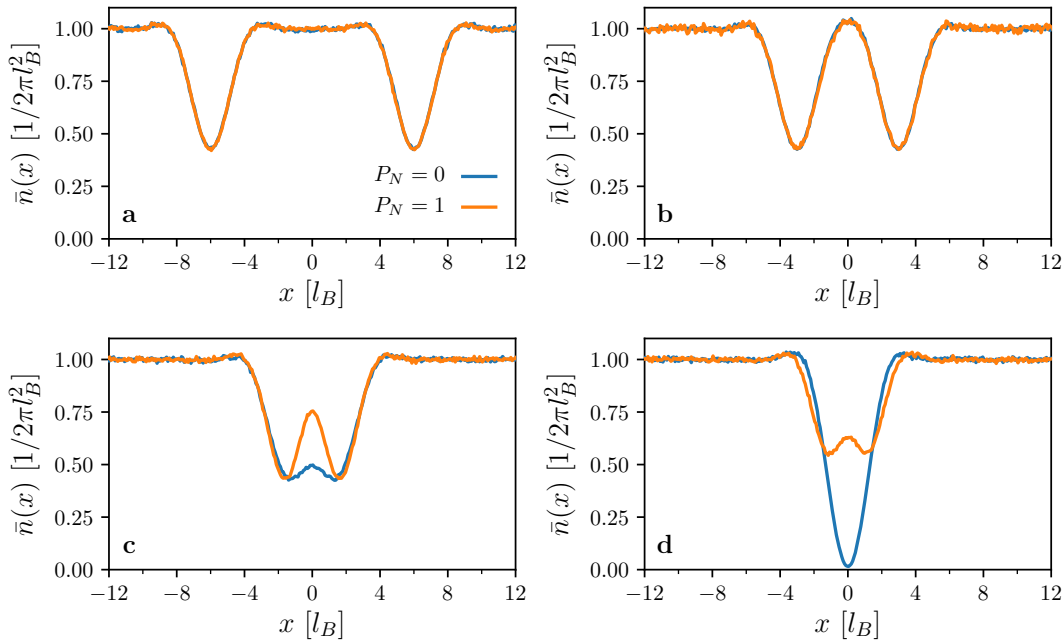

FIG. S3. **a-d**, Mean particle density $\bar{n}(x)$ along a cut connecting two $M = 1$ Moore-Read quasiholes located at $\eta_{1,2} = \pm R$, with $R/l_B = 6, 3, 1.5$ and $0$ for **a**, **b**, **c** and **d** respectively. For each value of $R$, the blue (orange) solid line indicates the $P_N = 0$ ($P_N = 1$) case. To obtain $\bar{n}(x)$, we average the density profile over a stripe with $|y| < 0.64\, l_B$.

The topological protection characterizing these states ensures that, once we create the QHs (far from each other and from the system boundary) in a given fusion channel, this will not change if we adiabatically bring the QHs close to each other (and eventually fuse them). On the one hand, this allowed us to study the behavior of the QH density depletions during the fusion process just through the Monte Carlo sampling of the wave functions (S6) and (S7) for different values of the QH coordinates [see Fig. S3]. On the other hand, it guarantees that, as long as one is able to create the MR QHs (far from each other) in the two different fusion channels, the experimental procedure we proposed in the main text can be applied to probe the non-Abelian nature of the MR QHs. In the main text we took advantage of this fact to overcome the difficulties in the creation of two overlapping QHs in the "odd" fusion channel [see orange line in Fig. S3 **d**]. Note that in the presence of an external potential the state in Eq. (S7) with

overlapping QHs may not be the ground state of the system, since it could be energetically more expensive than just exciting a fermionic edge mode on the boundary [13–15].

The question is whether or not one can create two spatially separated MR QHs in both fusion channels, and in particular in the "odd" one. To answer this question one should perform numerical experiments starting from a model Hamiltonian and a cooling mechanism, but this goes beyond the goal of the current work. What we do in the following instead, is trying to give strong arguments in favor our statement –namely that in an actual experiments the QH fusion channel is randomly chosen at each repetition. It is known from the literature [see Refs. [15, 16]] that in the presence of fermionic excitations, the one to one correspondence between the parity of $N$ and the fusion channel of the two MR QHs breaks down. To be precise, the fusion channel chosen by the QHs will be fixed by the parity of $(N - F)$, where $F$ is the number of fermionic edge excitations. At the same time, one can reasonably argue that the energy of the state with $F = 0$ and the one with $F = 1$, both hosting two well separated QHs, will be very similar (the energy difference between these two states will probably depend on both the distance between the QHs and the particular properties of the confining potential under study). Therefore, since in the cooling process the system does not necessarily reach the ground state (especially when there are almost-degenerate states), it is reasonable to expect that in an actual experiment the MR QHs will randomly select one of the two different fusion channels.

## VI. RIGID ROTATIONS OF THE MOORE-READ 4-QUASIHOLE STATES

In the following we consider the effect of rigid rotations on MR states with four QHs, where, for a given choice of the QH coordinates, two different ground states appear and the QHs show their actual non-Abelian behavior. We demonstrate that the non-Abelian terms of the QH braiding matrices only enter in the basis-change matrix $\mathcal{B}$, if we work with the PR basis. This proves that only rigid rotations of the anyons by angles $\theta_f \neq 2\pi k$ –with $k$ integer– can induce transformations which mix the different ground states. This observation is important in the context of theoretical studies, where, from the knowledge of the initial ground states, one can compute both $\mathcal{U}_B$ and $\mathcal{B}$ for any rigid rotation of the anyons.

The MR 4QH wave functions $\Psi_{0,1}(\eta; z)$ are obtained as the conformal blocks of suitable conformal field theories [8, 17, 18]:

$$\Psi_0(\eta; z) = \prod_{\mu<\nu}^{4} \eta_{\mu\nu}^{\frac{1}{4M}-\frac{1}{8}} \frac{(\eta_{13}\,\eta_{24})^{1/4}}{\sqrt{1+\sqrt{1-x}}} \left[ \Psi_{(13)(24)}(\eta; z) + \sqrt{1-x}\, \Psi_{(14)(23)}(\eta; z) \right] e^{-\frac{1}{8M}\sum_{\mu=1}^{4}|\eta_\mu|^2}, \quad (S36)$$

$$\Psi_1(\eta; z) = \prod_{\mu<\nu}^{4} \eta_{\mu\nu}^{\frac{1}{4M}-\frac{1}{8}} \frac{(\eta_{13}\,\eta_{24})^{1/4}}{\sqrt{1-\sqrt{1-x}}} \left[ \Psi_{(13)(24)}(\eta; z) - \sqrt{1-x}\, \Psi_{(14)(23)}(\eta; z) \right] e^{-\frac{1}{8M}\sum_{\mu=1}^{4}|\eta_\mu|^2}, \quad (S37)$$

where

$$\eta_{\mu\nu} = \eta_\mu - \eta_\nu, \qquad x = \frac{\eta_{12}\,\eta_{34}}{\eta_{13}\,\eta_{24}} = \frac{(\eta_1 - \eta_2)(\eta_3 - \eta_4)}{(\eta_1 - \eta_3)(\eta_2 - \eta_4)}, \quad (S38)$$

and

$$\Psi_{(ab)(cd)}(\eta; z) = \text{Pf}\left( \frac{(\eta_a - z_i)(\eta_b - z_i)(\eta_c - z_j)(\eta_d - z_j) + (i \leftrightarrow j)}{z_i - z_j} \right) \prod_{i<j}(z_i - z_j)^M e^{-\frac{1}{4}\sum_i^N |z_i|^2}. \quad (S39)$$

As a first step, we study what happens to the Pfaffian in equation (S39), when one rotates the particle coordinates by an angle $\theta$. We consider, in particular, the term

$$Q(\eta_a, \eta_b, \eta_c, \eta_d; z_k, z_j) \equiv \frac{(\eta_a - z_k)(\eta_b - z_k)(\eta_c - z_j)(\eta_d - z_j) + (\eta_a - z_j)(\eta_b - z_j)(\eta_c - z_k)(\eta_d - z_k)}{z_k - z_j}, \quad (S40)$$

which changes as

$$Q(\eta_a, \eta_b, \eta_c, \eta_d; z_k e^{-i\theta}, z_j e^{-i\theta}) = e^{-i3\theta}\, Q(\eta_a e^{i\theta}, \eta_b e^{i\theta}, \eta_c e^{i\theta}, \eta_d e^{i\theta}; z_k, z_j) \quad (S41)$$

when particle coordinates are rotated by an angle $\theta$.

In the explicit form of the Pfaffian appearing in the wave functions $\Psi_{(ab)(cd)}(\eta;z)$ we have a sum (with alternating signs) of terms given by the product of $N/2$ factors like (S40). Therefore, the effect of particle rotations on such wave functions is the following:

$$\begin{aligned}\hat{\mathcal{R}}_\theta\,\Psi_{(ab)(cd)}(\eta;z) &= \Psi_{(ab)(cd)}(\eta;ze^{-i\theta}) \\ &= e^{-i\frac{3N}{2}\theta}\,e^{-i\left[\frac{MN(N-1)}{2}\right]\theta}\,\Psi_{(ab)(cd)}(\eta e^{i\theta};z) \\ &= e^{-i\frac{3N}{2}\theta}\,e^{-i\left[\frac{MN(N-1)}{2}\right]\theta}\,\mathcal{S}_\theta\,\Psi_{(ab)(cd)}(\eta;z),\end{aligned} \tag{S42}$$

where the second term on the right-hand side comes from the Laughlin-like part of the wave functions $\Psi_{(ab)(cd)}(\eta;z)$. We recall that $\hat{\mathcal{R}}_\theta$ and $\mathcal{S}_\theta$ are the transformations mapping a generic wave function $\Psi(\eta;z)$ into $\Psi(\eta;ze^{-i\theta})$ and $\Psi(\eta e^{i\theta};z)$, respectively.

Finally, by noting that $x$ [see equation (S38)] is invariant under rotation of the QH coordinates, we can write:

$$\begin{aligned}\hat{\mathcal{R}}_\theta\,\Psi_{0,1}(\eta;z) &= \Psi_{0,1}(\eta;ze^{-i\theta}) \\ &= e^{-i\left[6\left(\frac{1}{4M}-\frac{1}{8}\right)+\frac{1}{2}\right]\theta}\,e^{-i\frac{3N}{2}\theta}\,e^{-i\left[\frac{MN(N-1)}{2}\right]\theta}\,\Psi_{0,1}(\eta e^{i\theta};z) \\ &= e^{-i\,\Sigma(N,M)\,\theta}\,\mathcal{S}_\theta\,\Psi_{0,1}(\eta;z),\end{aligned} \tag{S43}$$

where

$$\Sigma(N,M) \equiv \left[6\left(\frac{1}{4M}-\frac{1}{8}\right)+\frac{1}{2}\right] + \frac{3N}{2} + \left[\frac{MN(N-1)}{2}\right]. \tag{S44}$$

The first term on the right-hand side of the previous equation represents the Abelian part of the topological contribution to $\mathcal{U}(T)$. Note that it compensates the one coming from the $\eta$-depending normalization pre-factor in the wave functions $\Psi_{0,1}(\eta;z)$ when we apply $\mathcal{S}_\theta$ on them [see equations (S36) and (S37)], in agreement with the fact that in the PR basis such a term is contained in $\mathcal{U}_\mathrm{B}$. At the same time, the second and third terms represent the mean angular momentum $\langle\hat{L}_z\rangle_{0,0,0,0}$ of the 4QH state with all QHs in the origin and they cancel the same term entering – with the opposite sign– in $\mathcal{U}_\mathrm{B}$ [cf. equation (S28)]. As a consequence, the remaining non-topological contribution to the product $\mathcal{U}_\mathrm{B}\mathcal{B}$ is the Aharonov-Bohm phase due to the QH rotation.

Equation (S43) allows us to express the matrix $\mathcal{B}$ –written in the PR basis– in terms of the one written in the AR basis $\tilde{\mathcal{B}}$:

$$\begin{aligned}\mathcal{B}_{\alpha\beta}(\theta_f) &= \langle\Psi_\alpha(\eta;z)|\hat{\mathcal{R}}_{\theta_f}|\Psi_\beta(\eta;z)\rangle \\ &= e^{-i\,\Sigma(N,M)\theta_f}\,\langle\Psi_\alpha(\eta;z)|\mathcal{S}_{\theta_f}|\Psi_\beta(\eta;z)\rangle \\ &= e^{-i\,\Sigma(N,M)\theta_f}\,\tilde{\mathcal{B}}_{\alpha\beta}(\theta_f).\end{aligned} \tag{S45}$$

Since the unitary transformation $\mathcal{U}(T)$ does not depend on the basis choice, its non-Abelian contributions –coming only from the QH exchanges– must be the same if we consider $\mathcal{U}_\mathrm{B}\mathcal{B}$ or $\tilde{\mathcal{U}}_\mathrm{B}\tilde{\mathcal{B}}$. At the same time, in Ref. [8] it has been proved that, for an arbitrary closed path in the parameter space of QH coordinates (including rigid rotations of the QHs), all topological contributions to $\mathcal{U}(T)$ –both the Abelian and the non-Abelian ones– lie in the basis-change term $\tilde{\mathcal{B}}$, while $\tilde{\mathcal{U}}_\mathrm{B}$ corresponds to the Aharonov-Bohm phase due to the QH motion. Therefore, since equation (S45) relates $\mathcal{B}$ with $\tilde{\mathcal{B}}$, we can conclude that all non-Abelian contributions to $\mathcal{U}(T)$ lie in the basis-change matrix $\mathcal{B}$ also if we choose the PR basis.

---

[1] L. E. Picasso, *Lectures in Quantum Mechanics* (Springer International Publishing, 2016).
[2] M. S. Baraban, *Low Energy Excitations in Quantum Condensates*, Ph.D. thesis (2010).
[3] N. Metropolis, A. W. Rosenbluth, M. N. Rosenbluth, A. H. Teller, and E. Teller, "Equation of State Calculations by Fast Computing Machines," The Journal of Chemical Physics **21**, 1087–1092 (1953).
[4] W. Krauth, *Statistical Mechanics: Algorithms and Computations* (Oxford University Press, 2006).
[5] R. Morf and B. I. Halperin, "Monte Carlo evaluation of trial wave functions for the fractional quantized Hall effect: Disk geometry," Phys. Rev. B **33**, 2221–2246 (1986).
[6] H. Kjønsberg and J. Myrheim, "Numerical study of charge and statistics of Laughlin quasiparticles," International Journal of Modern Physics A **14**, 537–557 (1999).

[7] R. O. Umucalılar, E. Macaluso, T. Comparin, and I. Carusotto, "Time-of-Flight Measurements as a Possible Method to Observe Anyonic Statistics," Phys. Rev. Lett. **120**, 230403 (2018).
[8] P. Bonderson, V. Gurarie, and C. Nayak, "Plasma analogy and non-Abelian statistics for Ising-type quantum Hall states," Phys. Rev. B **83**, 075303 (2011).
[9] M. Born and V. Fock, "Beweis des Adiabatensatzes," Zeitschrift für Physik **51**, 165–180 (1928).
[10] G. Rigolin and G. Ortiz, "Adiabatic theorem for quantum systems with spectral degeneracy," Phys. Rev. A **85**, 062111 (2012).
[11] F. Wilczek and A. Zee, "Appearance of Gauge Structure in Simple Dynamical Systems," Phys. Rev. Lett. **52**, 2111–2114 (1984).
[12] Y. Aharonov and D. Bohm, "Significance of Electromagnetic Potentials in the Quantum Theory," Phys. Rev. **115**, 485–491 (1959).
[13] X.-G. Wen, "Topological orders and edge excitations in fractional quantum Hall states," Advances in Physics **44**, 405–473 (1995), https://doi.org/10.1080/00018739500101566.
[14] M. Milovanović and N. Read, "Edge excitations of paired fractional quantum Hall states," Phys. Rev. B **53**, 13559–13582 (1996).
[15] X. Wan, K. Yang, and E. H. Rezayi, "Edge Excitations and Non-Abelian Statistics in the Moore-Read State: A Numerical Study in the Presence of Coulomb Interaction and Edge Confinement," Phys. Rev. Lett. **97**, 256804 (2006).
[16] X. Wan, Z.-X. Hu, E. H. Rezayi, and K. Yang, "Fractional quantum Hall effect at $\nu = 5/2$: Ground states, non-Abelian quasiholes, and edge modes in a microscopic model," Phys. Rev. B **77**, 165316 (2008).
[17] G. Moore and N. Read, "Nonabelions in the fractional quantum Hall effect," Nuclear Physics B **360**, 362 – 396 (1991).
[18] C. Nayak and F. Wilczek, "$2n$-quasihole states realize $2^{n-1}$-dimensional spinor braiding statistics in paired quantum Hall states," Nuclear Physics B **479**, 529 – 553 (1996).




\end{document}